\documentclass[journal]{IEEEtran}
\usepackage{xcolor,soul,framed} 
\usepackage{cite}
\usepackage{amsmath,amssymb,amsfonts}
\usepackage{mathtools}
\usepackage{graphicx}
\graphicspath{ {./plots/} }
\usepackage{textcomp}
\usepackage{dcolumn}
\usepackage{bm}
\usepackage{subfigure} 
\usepackage{pict2e}
\usepackage{epstopdf}
\usepackage{arydshln}
\usepackage{gensymb}
\usepackage{lipsum}

\newcommand{\Fig}[1]{Fig.\,{\ref{#1}}}
\newcommand{\Figs}[1]{Figs.\,{\ref{#1}}}

\newcommand{\jj}{\mathrm{j}}
\newcommand{\cp}{\text{cp}}
\newcommand{\xp}{\text{xp}}
\newcommand{\cpL}{\text{cp,(L)}}
\newcommand{\xpL}{\text{xp,(L)}}
\newcommand{\cpR}{\text{cp,(R)}}
\newcommand{\xpR}{\text{xp,(R)}}

\newcommand{\sump}{\sideset{}{'}\sum}

\newcommand{\Eap}{\boldsymbol{\mathcal{E}}}
\newcommand{\Jpa}{\boldsymbol{\mathcal{J}}}

\begin{document}

\onecolumn
\newpage
\thispagestyle{empty}

\textbf{Copyright:}
\copyright 2021 IEEE. Personal use of this material is permitted.  Permission from IEEE must be obtained for all other uses, in any current or future media, including reprinting/republishing this material for advertising or promotional purposes, creating new collective works, for resale or redistribution to servers or lists, or reuse of any copyrighted component of this work in other works.\\

\textbf{Disclaimer:} This work has been published in \textit{IEEE Access}. \\

Citation information: 10.1109/ACCESS.2021.3096715

\newpage

\twocolumn

\title{Cross-polarization Control in FSSs by means of an Equivalent Circuit Approach}


\author{ Carlos Molero, Antonio Alex-Amor,  Francisco Mesa, Angel Palomares-Caballero, and Pablo Padilla 
\thanks{This work was supported by the Spanish Research and Development National Program under Projects TIN2016-75097-P, RTI2018-102002-A-I00, B-TIC-402-UGR18, TEC2017-84724-P, and the predoctoral grant FPU18/01965; by Junta de Andalucía under project P18-RT-4830.}
\thanks{C. Molero, A. Alex-Amor, A. Palomares-Caballero, and P. Padilla are with the Departamento de Teor\'{i}a de la Se\~{n}al, Telem\'{a}tica y Comunicaciones, Universidad de Granada, 18071 Granada, Spain (email: cmoleroj@ugr.es; aalex@ugr.es; angelpc@ugr.es; pablopadilla@ugr.es)}
\thanks{F. Mesa is with the Microwaves Group, Department of Applied Physics 1, Escuela Técnica Superior de Ingenieria Informática, Universidad de Sevilla, 41012 Sevilla, Spain;  (e-mail: mesa@us.es)}
}

\markboth{}%
{Molero \MakeLowercase{\textit{et al.}}:  Cross-polarization Control in FSS by means of an Equivalent Circuit Approach}

\maketitle

\newcommand*{\bigs}[1]{\vcenter{\hbox{\scalebox{2}[8.2]{\ensuremath#1}}}}

\newcommand*{\bigstwo}[1]{\vcenter{\hbox{ \scalebox{1}[4.4]{\ensuremath#1}}}}

\begin{abstract}
This paper presents an efficient equivalent circuit approach (ECA), based on a Floquet modal expansion, for the study of the co- and cross-polarization in frequency selective surfaces (FSS) formed by periodic arrays of patches/apertures in either single or stacked configurations. The ECA makes it possible the derivation of analytical expressions for the generalized scattering parameters associated with the proposed circuit networks. Furthermore, the proposed circuit approach is an efficient surrogate model that can be combined with optimization techniques and artificial intelligence algorithms for the efficient design of FSS structures, saving efforts in the computation compared to time-consuming full-wave simulators and tedious synthesis (simulation-assisted) techniques. Due to the simplicity of the topology of the involved networks, the ECA can also be advantageously used to gain physical insight. The proposed approach is applied and validated in different FSS configurations where the cross-pol component plays a fundamental role in the design, as in circular polarizers, polarization rotators, and reflectarray cells.
\end{abstract}



\IEEEpeerreviewmaketitle

\section{Introduction}

\IEEEPARstart{C}{ontrolling} the behavior of the electromagnetic (EM) waves is a traditional challenge of the microwave and antenna engineering community~\cite{Macfarlane1946, Felsen1973}, which still raises a lot of attention~\cite{RoadmapMetasurfaces, Li2020}. The modification of the transmission and reflection responses of an impinging EM wave on a periodic structure has been studied for different frequency ranges, with some examples from microwave to terahertz regimes found in~\cite{microwaves3, microwaves4, Microwave_strips, Microwave_apertures, Qu2020, THz_strips, THz_aperture, SJLi2020, SJLi2021}. In these works, the selected structure to provide the desired control of the incident EM is the so-called frequency selective surface (FSS)~\cite{Munk2000, FSS2}. The FSS can be classified into two categories depending of the implemented geometry of the unit cell; namely, FSS based on either apertures or patches~\cite{Munk2000}. 

One of the most appreciated functionalities in the design of FSS is the control of the cross-polarization. This capability enables the design of polarization-converter devices~\cite{Science_polarizer, Optical_polarizer} because both the co-polarized and cross-polarized components of the EM wave have to be tuned in order to obtain the desired polarization (horizontal, vertical, circular or elliptical). Polarization converters based on apertures~\cite{Rotator_aperture_1, Rotator_aperture_2, Rotator_aperture_3, Rotator_aperture_4, clendinning} and based on patches \cite{Rotator_patch_1,Rotator_patch_2,Page-AP2020} are reported in the literature. In most of the previous works, the incident wave is transmitted along the FSS structure, although there are also FSSs where the incident wave is totally reflected. This fact is taken into advantage in radar cross section (RCS) applications in order to rotate the polarization of the incoming wave towards its orthogonal polarization and, thus, achieve a RCS reduction~\cite{RCS_1, RCS_2, RCS_3, RCS_4}. 

Given the importance of controlling the cross polarization in transmission and/or in reflection by means of FSS, an efficient design of these structures is imperative. The optimization of the FSS design can be carried out by using an EM simulator. However, its computational cost is very high because hundreds of full-wave simulations are generally needed to reach the desired goals. An alternative strategy to substantially reduce the computational cost is replacing the full-wave simulations by equivalent circuit models that represent the performance of the FSS in a cost-effective way~\cite{ec8, ec9, Page-AP2018, costa_magazine, eca_magazine, heuristic1}. These circuit models can then be combined with optimization and artificial intelligence algorithms for an efficient design of the device, specially when a large number of iterations are required in the process. Among the different techniques to obtain a circuit model of periodic structures~\cite{heuristic2, Page-AP2020, Costa-AP2020, glide_qiao, glide_circuitmodel, Al-Gburi2020,  Guo2018}, the equivalent circuit models based on Floquet analysis \cite{fss_eca2006, Berral2015, molero_symmetrical1D, evenodd1} have shown remarkable outcomes even for intricate structures~\cite{Berral2015, Hum2017, arbitrary2D_frezza, Carlos_MTT3D, multimodalTAP, Hum2021}. One of the most relevant characteristics of this method comes from its analytical (or quasi-analytical) nature. It means that no previous full-wave frequency-sweeping simulations are needed in order to accurately take into account the effects of higher order harmonics and couplings in single and stacked FSS structures in a very extended frequency band~\cite{fss_eca2006, Berral2015, multimodalTAP}.

In this work, we present analytical expressions for both co-pol and cross-pol scattering parameters of single and stacked aperture- and patched-based FSSs. In order to have complete independence of full-wave simulators, the formulation is based on Floquet modes in contrast with other state-of-the-art synthesis techniques~\cite{Costa-AP2020, Page-AP2020, Page-AP2018}. 
Furthermore, as discussed in our previous work \cite{multimodalTAP}, the analytical nature of the approach makes that linear transformations of the scatterers (such as displacement, rotation, and/or scaling) can easily be introduced to model in a simple and efficient manner complex FSS stacks.  The present work goes one step further in that direction and includes the treatment of the cross-pol component in both aperture- and patch-based FSS structures. This provides a great advantage by allowing the aperture/patch scatterers can be different between layers. It represents an improvement with respect to~\cite{Hum2021}, where only patch-based FSSs with the same base geometry are studied.

The paper is organized as follows. Section~\ref{sec:Theory} presents the derivations and the analytical expressions for the co- and cross-polarization in aperture-based FSS. In Section~\ref{Patch-arrays}, the calculation of the generalized scattering parameters for patched FSS is given. Both sections discuss the application of the formulation for both single and stacked FSS. Section~\ref{Numerical examples} shows several examples of FSS based on either apertures or patches that are characterized using the developed analytical expressions. Finally, in Section~\ref{sec:Conc}, the main conclusions of the work are drawn.

\section{ Cross-polarization in aperture arrays}
\label{sec:Theory}

\subsection{Single aperture problem}
\begin{figure}[t]
\begin{center}
\subfigure[]{
\includegraphics[width=0.44\columnwidth]{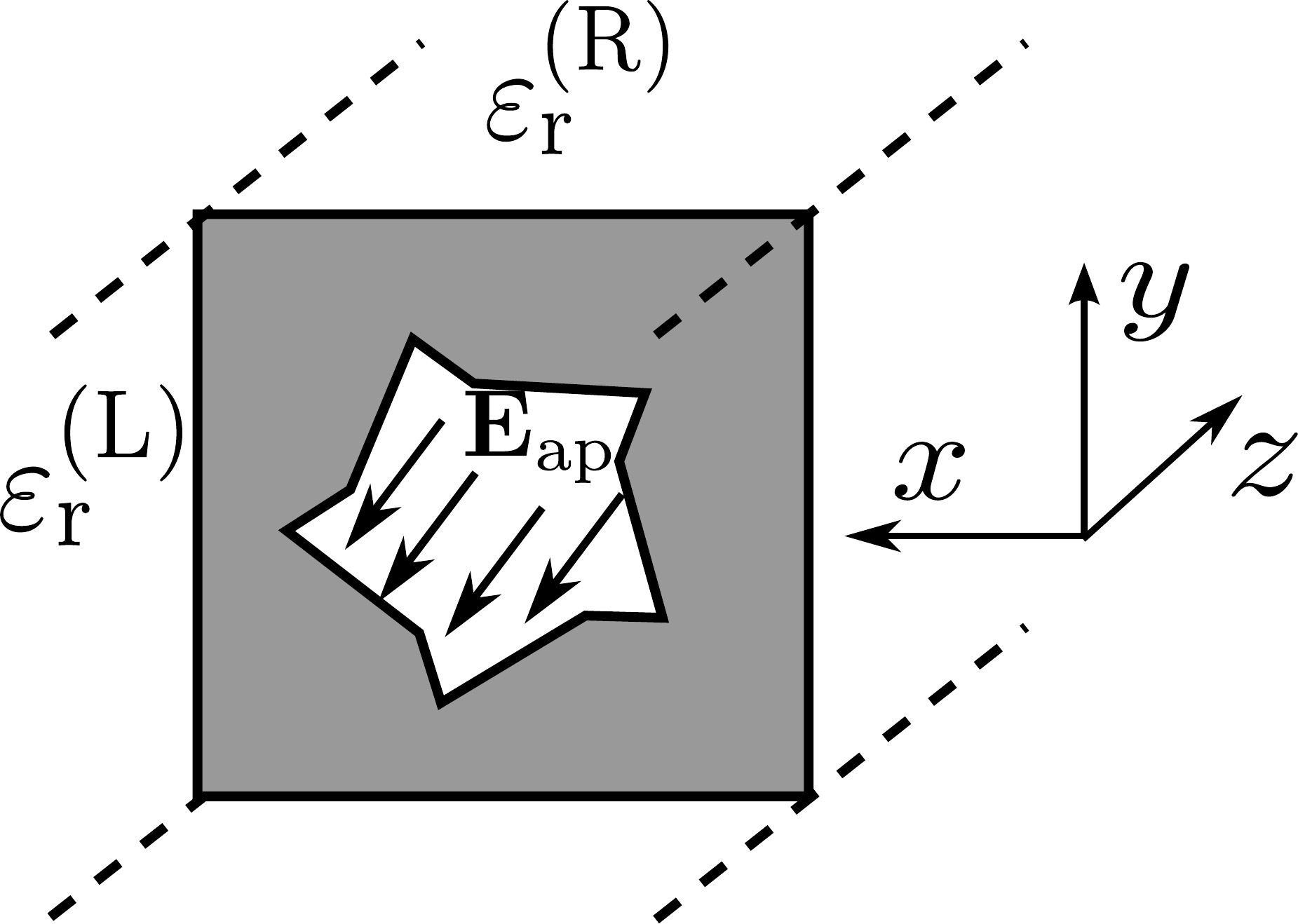}} %
\subfigure[]{
\includegraphics[width=0.52\columnwidth]{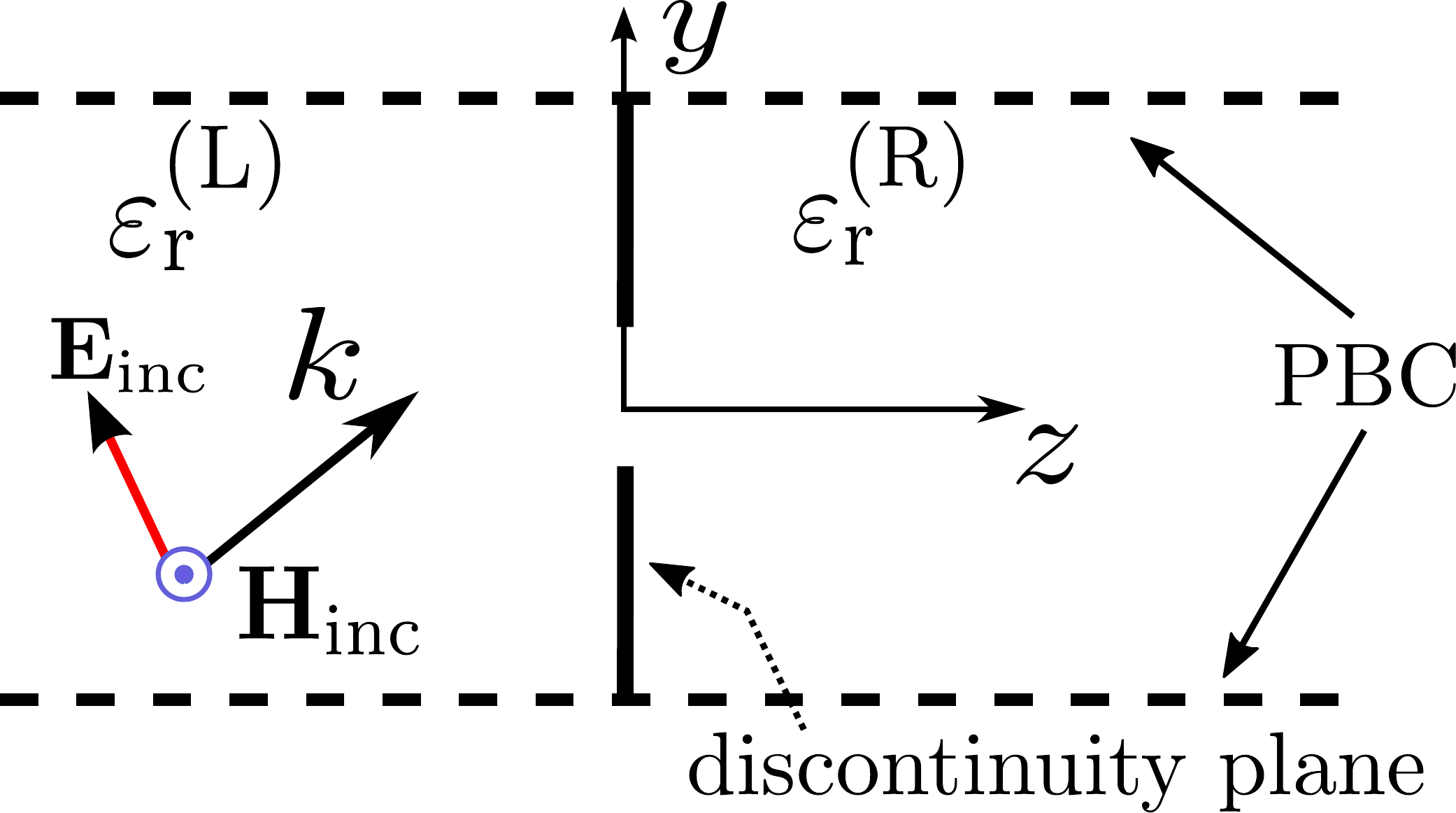}}
\end{center}
\caption{\label{single_array} (a): Unit cell in perspective. (b): Unit cell in the $yz$-plane.} 
\end{figure}

The unit cell for the problem under consideration in the present section is depicted in~\Fig{single_array}, where the scattering of an impinging time-harmonic ($\mathrm{e}^{\jj \omega t}$) plane wave on an infinite 2-D periodic structure has been reduced to a classical discontinuity problem in waveguides. For simplicity, we will consider but suppress the $\mathrm{e}^{\jj \omega t}$ dependence in the following derivation. The planar discontinuity consists of a infinitesimally thin metallic sheet  with an arbitrary-shaped aperture. This discontinuity is located at~$z=0$ and embedded between two semi-infinite homogeneous and isotropic dielectric media with permittivity $\varepsilon_{\text{r}}^{(\text{L/R})}$, as shown in~\Fig{single_array}(b). The unit cell is bounded by periodic boundary conditions (PBC) and the incident plane wave is arbitrarily oriented. The electric field magnitude of the incident wave is assumed to be unity, thus letting its magnetic-field magnitude identical to the value of admittance associated with the wave. 

Next we follow the general procedure reported in~\cite{fss_eca2006, Berral2015} although here the necessary derivations to obtain closed-form expressions associated with the cross-pol components will be explicitly shown. Since the interaction of the incident wave with the discontinuity excites an infinite number of harmonics/modes in the unit cell, the \textit{transverse} electric field at both sides of the discontinuity plane can be described in terms of the following Floquet expansion:
\begin{multline}\label{eq:EL}
\mathbf{E}_t^{(\text{L})}(x, y; \omega) = (1 + R)\mathbf{e}_{00}^{\cp}(x, y) + V_{00}^{\xpL}\mathbf{e}_{00}^{\xp}(x, y) \\
+ \displaystyle\sump_{n, m} V_{nm}^{\cpL} \mathbf{e}_{nm}^{\cp}(x, y)  + 
 \displaystyle\sump_{n,m} V_{nm}^{\xpL} \mathbf{e}_{nm}^{\xp}(x, y)  
\end{multline} 
\vspace*{-6mm}
\begin{multline}\label{eq:ER}
\mathbf{E}_t^{(\text{R})}(x, y; \omega) = T\mathbf{e}_{00}^{\cp}(x, y) + V_{00}^{\xpR}\mathbf{e}_{00}^{\xp}(x, y) \\ 
+ \displaystyle\sump_{n, m} V_{nm}^{\cpR} \mathbf{e}_{nm}^{\cp}(x, y)  + \displaystyle\sump_{n, m} V_{nm}^{\xpR} \mathbf{e}_{nm}^{\xp}(x, y)  
\end{multline}
where the superscripts ``cp'' and ``xp'' denotes co- and cross-polarization terms with respect to the polarization of the incident wave, respectively. For TE incidence, the cross-polarization superscript becomes TM and vice versa. The superscript ``L/R'' denotes the left/right-side dielectric medium.  The factor~$1$ in~\eqref{eq:EL} accounts for the unit-amplitude of the incident electric field, here characterized by the harmonic of order $m,n =0,0$. The coefficients~$R$ and~$T$ are the corresponding copolar reflection and transmission coefficients, respectively. The coefficient $V_{00}^{\xp,\text{(L/R)}}$ is the amplitude of the 'cross-pol' harmonic of order $m,n= 0,0$ at both sides of the discontinuity, excited thanks to the interaction between the incident harmonic and the discontinuity. The coefficients $V_{nm}^{\cp/\xp, (\text{L})/(\text{R})}$ are the amplitudes associated with harmonics of order $n,m$ at both sides of the discontinuity as well. All the above coefficients are function of the  angular frequency ($\omega$) of the impinging wave. The prime symbol in the series stands for summation $\forall n,m$ except $n,m=0,0$. 

The spatial profile of the transverse electric field associated with each harmonic is represented by $\mathbf{e}_{nm}^{\cp/\xp}$ and defined in terms of TM/TE harmonics as
\begin{align}
    \mathbf{e}_{nm}^{\text{TM}}(x,y) &=  \frac{1}{\sqrt{p_{\text{x}} p_{\text{y}}}}\frac{\exp [-\mathrm{j}(k_{xn}x + k_{ym}y)]}{\sqrt{p_{x}p_{y}}}\,
    \hat{\mathbf{e}}_{nm}^{\text{TM}} \\
    \mathbf{e}_{nm}^{\text{TE}}(x,y) &= \frac{1}{\sqrt{p_{\text{x}} p_{\text{y}}}} \frac{\exp [-\mathrm{j}(k_{xn}x + k_{ym}y)]}{\sqrt{p_{x}p_{y}}}\,
    \hat{\mathbf{e}}_{nm}^{\text{TE}}
\end{align}
with 
\begin{align}
    \hat{\mathbf{e}}_{nm}^{\text{TM}} &= \frac{k_{xn} \mathbf{\hat{x}} + k_{ym} \mathbf{\hat{y}}}{\sqrt{k_{xn}^{2} + k_{ym}^{2}}}  \\
    \hat{\mathbf{e}}_{nm}^{\text{TE}} &= \frac{k_{ym} \mathbf{\hat{x}} - k_{xn} \mathbf{\hat{y}}}{\sqrt{k_{xn}^{2} + k_{ym}^{2}}} 
    \left[=\hat{\mathbf{e}}_{nm}^{\text{TM}}\times\mathbf{\hat{z}}\right] \\
    k_{xn} &= \sqrt{\varepsilon_{\text{r}}^{\text{(L)}}}k_{0}\sin\theta\cos\phi + \frac{2n\pi}{p_{x}} \\
    k_{ym} &= \sqrt{\varepsilon_{\text{r}}^{(\text{L})}}k_{0}\sin\theta\sin\phi + \frac{2m\pi}{p_{y}} \,
\end{align}
where $p_{\text{x}}$ and $p_{\text{y}}$ are the periodicity along $x$ and $y$ axis, $k_{0}$ the free-space wavenumber, and $\theta, \phi$ are the elevation and azimuth incident angles. 
At the discontinuity plane, the transverse electric field is different from zero only at the aperture region $\Omega$. A key assumption of the present approach is that the spatial profile, $\Eap$, of this transverse electric field at the aperture is considered to be frequency-independent; namely, the total transverse field, $\mathbf{E}_\text{ap}$, can be factorized in the following way
\begin{equation}
\mathbf{E}_{\text{ap}}(x,y;\omega) = A(\omega)\boldsymbol{\mathcal{E}}(x, y) \hspace{5 mm} \forall x, y \in \Omega   
\end{equation}
where $A(\omega)$ can be regarded as the amplitude which does depend on frequency whereas $\boldsymbol{\mathcal{E}}(x, y)$ describes the spatial distribution which does not depend on frequency. This approximation has been found to be valid up to the second excitable resonance of the considered scatterer. Nonetheless, this usually leads to a large operation bandwidth for the ECA, even within the grating-lobe regime \cite{eca_magazine, multimodalTAP}. In fact, the onset of the grating-lobe regime corresponds to the cutoff frequency of high-order propagating harmonics. In this sense, the ECA is capable of reproducing very exotic behaviors where several diffraction orders are present, regardless of the electrical size of the unit cell under the assumptions described above.

Since the electric field is continuous in the aperture region, we can write
\begin{equation}
    \mathbf{E}_t^{(1)}(x, y; \omega) = \mathbf{E}_t^{(2)} (x, y; \omega) \equiv \mathbf{E}_{\text{ap}} \hspace{5mm} \forall x, y \in \Omega\,.  
\end{equation}
If this continuity of the electric field is forced for the projection of the fields onto the modal functions $\mathbf{e}_{nm}^\text{TM/TE}$, it allows us to calculate the amplitudes of the harmonics $V_{nm}^{\text{cp/xp, (L/R)}}$ by means of the following expressions: 
\begin{align}
\label{eq:V00cp}
(1 + R) &= A(\omega)  \widetilde{\Eap}(k_{x0},k_{y0})\cdot  \hat{\mathbf{e}}_{00}^{\text{cp}} \\
\label{eq:V00xp}
V_{00}^{\xp,\text{(L/R)}} &= A(\omega) \widetilde{\Eap}(k_{x0},k_{y0}) \cdot \hat{\mathbf{e}}_{00}^{\text{xp}} \\
\label{eq:Vnm}
V_{nm}^{\cp/\xp,\text{(L/R)}} &= A(\omega)\widetilde{\Eap}(k_{xn},k_{ym}) \cdot \hat{\mathbf{e}}_{nm}^{\text{cp/xp}} \;.
\end{align}
 It should be noted that the expressions for $V_{nm}^{\text{cp/xp,(L/R)}}$ at the left- and right-hand side coincide, which allows us to remove the superscript (L/R) in the following. In the above expressions we have made use of the fact that
\begin{equation}
    \int_{\Omega} \Eap(x, y) \cdot [\mathbf{e}_{nm}^{\cp/\xp}(x, y)]^{*} \text{d}\Omega = \widetilde{\Eap}(k_{xn},k_{ym})\cdot 
    \hat{\mathbf{e}}_{nm}^{\text{cp/xp}}
\end{equation}
where 
\begin{equation}
    \widetilde{\Eap}(k_{xn},k_{ym}) =  \frac{1}{\sqrt{p_{x}p_{y}}}
    \int_\Omega  \Eap(x, y)\, \mathrm{e}^{\mathrm{j}(k_{xn}x + k_{ym}y)}  \,\text{d}\Omega
\end{equation}
can be recognized as the 2-D Fourier transform of the spatial profile $\Eap(x,y)$ at $(k_{xn},k_{ym})$. Thus, if we now define a frequency-independent parameter~$N_{nm}^{\cp/\xp}$ given by
\begin{equation} \label{eq:transformers}
N_{nm}^{\cp/\xp} = \widetilde{\Eap}(k_{xn},k_{ym})\cdot 
    \hat{\mathbf{e}}_{nm}^{\text{cp/xp}} 
\end{equation}
the expressions \eqref{eq:V00cp}\,--\,\eqref{eq:Vnm} can be rewritten in general as
\begin{equation}\label{VnmcoA} 
    \frac{V_{nm}^{\cp/\xp}}{N_{nm}^{\cp/\xp}} = 
    \frac{(1 + R)}{N_{00}^{\cp}}
\end{equation}
what indicates that each amplitude is actually proportional to the voltage factor $(1 + R)$. As already discussed in~\cite{Berral2015}, the factors relating each amplitude and $(1 + R)$ can be interpreted, from a circuit point of view, as the value of the turns ratio corresponding to transformers. It is worth noting here that these transformer ratios turn out to be be frequency independent, which will be very relevant from an operational standpoint.

\begin{figure}
\begin{center}
\subfigure[]{
\includegraphics[width=0.7\columnwidth]{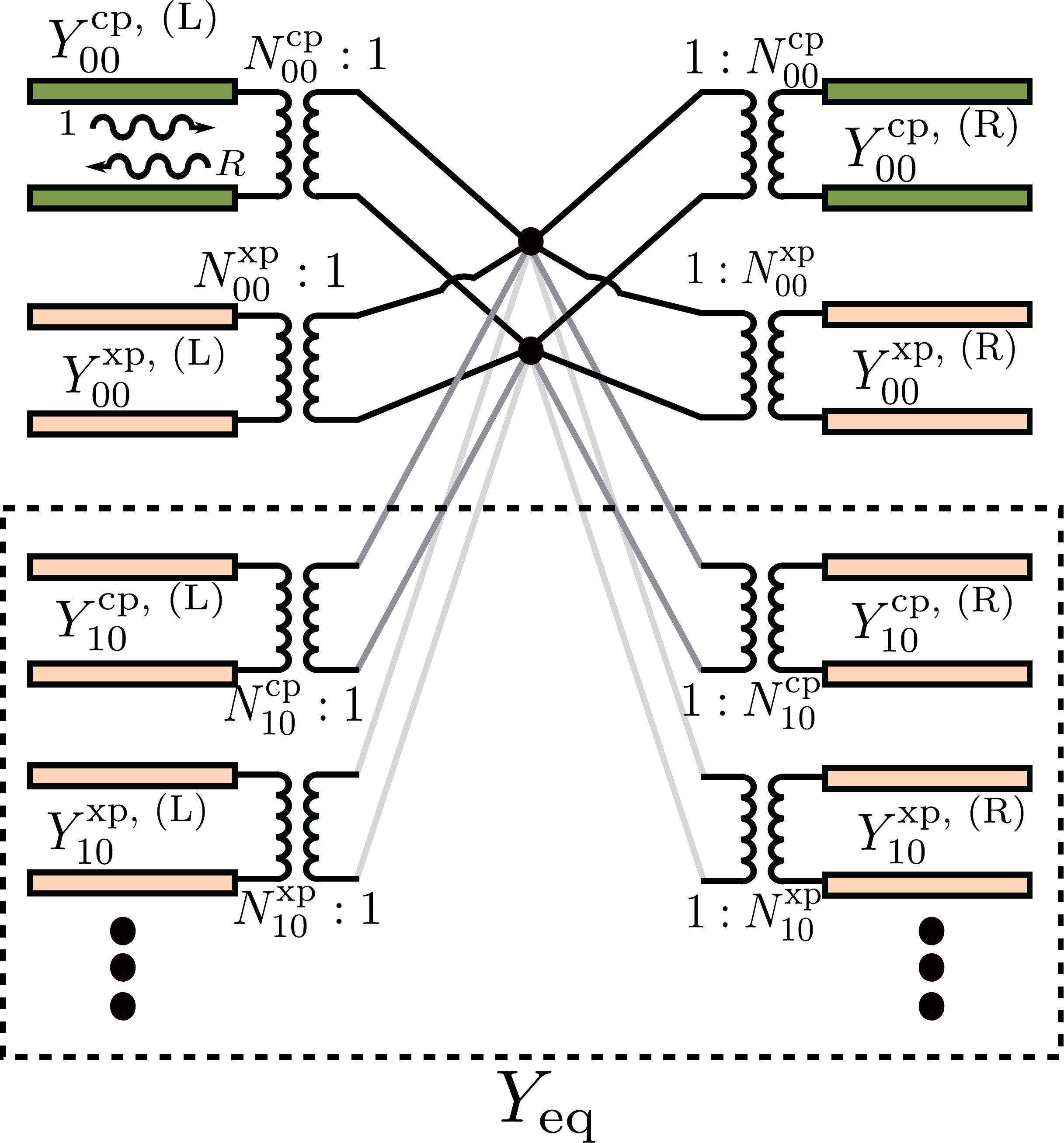} }
\subfigure[]{
\includegraphics[width=0.85\columnwidth]{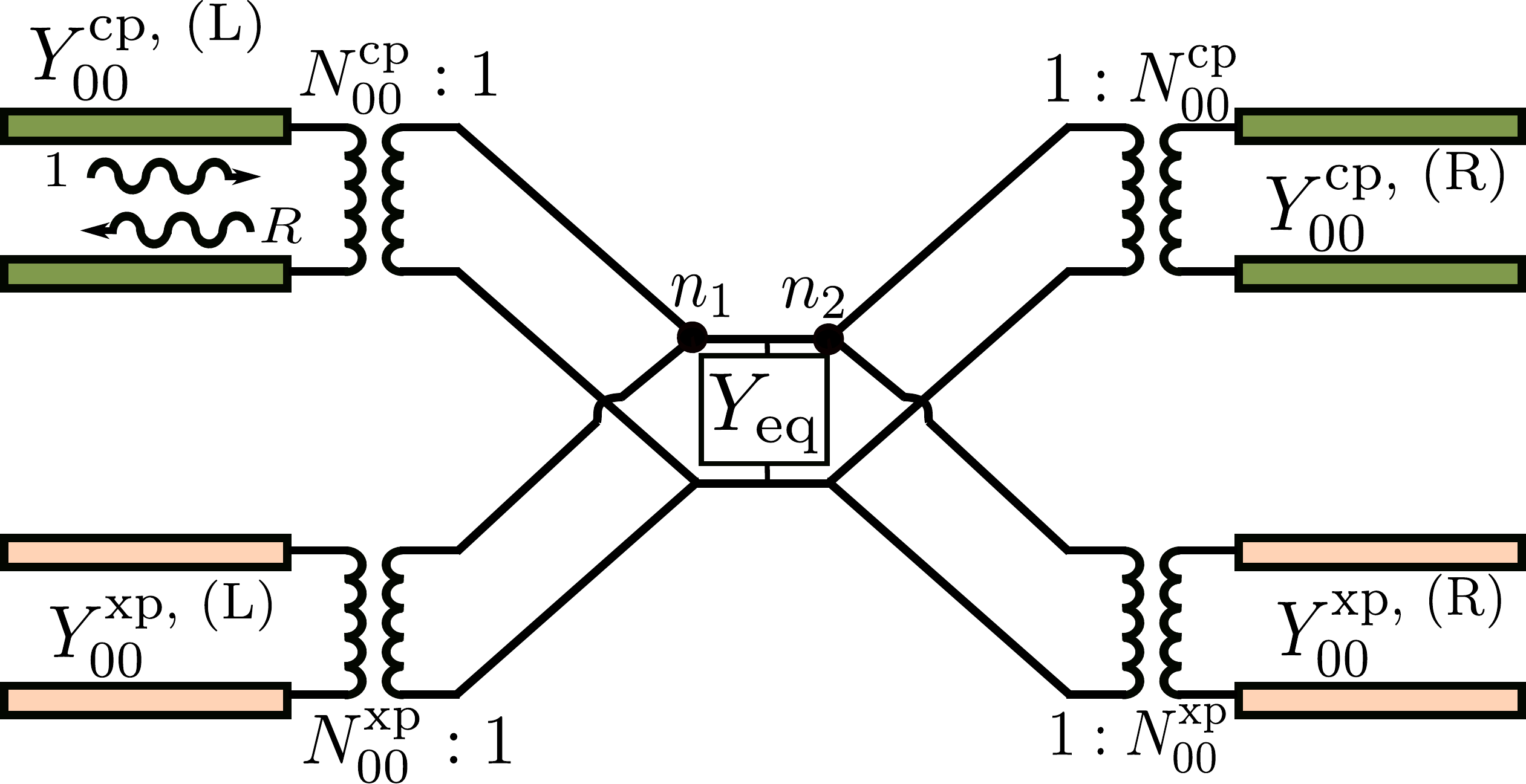} }
\end{center}
\caption{\label{circuito} (a): Multi-modal equivalent circuit for a single 2-D slot-like periodic array when assuming the incidence of the harmonic $n,m=0,0$. The interaction of the incident harmonic and the discontinuity excites all the higher-order harmonics, represented by their corresponding transmission lines. (b): Same equivalent circuit, with the higher-order harmonics grouped in the term $Y_{\text{eq}}$.} 

\end{figure}

If we now force the continuity of the power through the aperture:
\begin{equation}\label{Power}
   \displaystyle\int _{\Omega} \mathbf{E}_{\text{ap}}   \times [\mathbf{H}_t^{(\text{L})}]^{*} \text{d}\Omega = \displaystyle\int _{\Omega} \mathbf{E}_{\text{ap}}   \times [\mathbf{H}_t^{(\text{R})}]^{*} \text{d}\Omega
\end{equation}
it is then possible to obtain the reflection coefficient $R$ after considering that the transverse magnetic fields at both sides of the discontinuities are given by the following Floquet expansions:
\begin{multline}\label{H1}
\mathbf{H}_t^{(\text{L})}(x, y; \omega) \\ = (1 - R)Y_{00}^{\cpL} \mathbf{h}_{00}^{\cp}(x, y)  - V_{00}^{\xp}Y_{00}^{\xpL} \mathbf{h}_{00}^{\xp}(x, y)  \\  - \displaystyle\sump_{n, m} V_{nm}^{\cp} Y_{nm}^{\cpL} \mathbf{h}_{nm}^{\cp}(x, y) - \displaystyle\sump_{n, m} V_{nm}^{\xp} Y_{nm}^{\xpL} \mathbf{h}_{nm}^{\xp}(x, y) 
\end{multline} 
\vspace*{-4mm}
\begin{multline}\label{H2}
\mathbf{H}_t^{(\text{R})}(x, y; \omega) \\ = TY_{00}^{\cpR} \mathbf{h}_{00}^{\cp}(x, y)  + V_{00}^{\xp}Y_{00}^{\xpR}\mathbf{h}_{00}^{\xp}(x, y)  \\ + \displaystyle\sump_{n, m} V_{nm}^{\cp} Y_{nm}^{\cpR}\mathbf{h}_{nm}^{\cp}(x, y) + \displaystyle\sump_{n, m} V_{nm}^{\xp} Y_{nm}^{\xpR} \mathbf{h}_{nm}^{\xp}(x, y)
\end{multline}
where $\mathbf{h}_{nm}^{\cp/\xp}(x, y) = \mathbf{\hat{z}} \times \mathbf{e}_{nm}^{\cp/\xp}(x, y)$ and $Y_{nm}^{\text{cp/xp,(L/R)}}$ is the admittance associated with the harmonics of order~$(n, m)$:
\begin{align}
Y_{nm}^{\text{TE,(L/R)}} &= \frac{\beta_{nm}^{\text{(L/R)}}}{\omega \mu_{0}}    \\
Y_{nm}^{\text{TM,(L/R)}} &= \frac{\omega \varepsilon_{0}\varepsilon_{\text{r}}^{\text{(L/R)}}}{\beta_{nm}^{\text{(L/R)}}} 
\end{align}
with the longitudinal wavenumber, $\beta_{nm}^{\text{(L/R)}}$, given by
\begin{equation}
\beta_{nm}^{\text{(L/R)}} = \sqrt{\varepsilon_{\text{r}}^{\text{(L/R)}} k_{0}^{2} - k_{xn}^2 - k_{ym}^{2}}\;.
\end{equation}
Thus, introducing \eqref{H1} and \eqref{H2} into \eqref{Power}, the reflection coefficient is found to be 
\begin{equation}\label{Raperture}
    R = \frac{|N_{00}^{\text{cp}}|^{2}(Y_{00}^{\cpL} - Y_{00}^{\cpR}) - |N_{00}^{\xp}|^{2} (Y_{00}^{\xpL} + Y_{00}^{\xpR} ) - Y_{\text{eq}}}{|N_{00}^{\text{cp}}|^{2}(Y_{00}^{\cpL} + Y_{00}^{\cpR}) + |N_{00}^{\xp}|^{2} (Y_{00}^{\xpL} + Y_{00}^{\xpR} ) + Y_{\text{eq}}}
\end{equation}
with
\begin{multline}
Y_{\text{eq}} = \displaystyle\sump_{nm} |N_{nm}^{\cp}|^{2} (Y_{nm}^{\cpL} + Y_{nm}^{\cpR}) \\ 
+ \displaystyle\sump_{n, m} |N_{nm}^{\xp}|^{2}(Y_{nm}^{\xpL} + Y_{nm}^{\xpR})\,.
\end{multline}
Notice that $Y_{\text{eq}}$ corresponds to an infinite summation of admittances connected in parallel. Each individual admittance, $Y_{nm}^{\text{cp/xp,(L/R)}}$, is associated with a semi-infinite transmission line section that is connected to the input line (corresponding to the incident harmonic) through a transformer of turns ratio~$N_{nm}^{\cp/\xp}$. Electromagnetic coupling between adjacent cells is taken into consideration by the \textit{multi-modal} nature of the present equivalent approach; namely, Floquet harmonics of lower and higher order are coupled together through the parallel connections appreciated in \Fig{circuito}.

For the sake of simplicity, $Y_{\text{eq}}$ will be considered as a single admittance (it includes the contribution of all the harmonics except the ones corresponding to $m, n = 0,0$). The corresponding 4-port equivalent circuit can therefore be schematically represented as the one in~\Fig{circuito}. Four transmission lines connected in parallel are distinguished: two lines in green that describe the propagation of the incident harmonic through the media (L/R) and two lines in pink that describe the propagation of the $(0,0)$ cross-polarized harmonic in both dielectric media. All these lines are connected to~$Y_{\text{eq}}$ via transformers with turns ratio~$N_{00}^{\text{cp}/\text{xp}}$. 

Standard circuit rules can now be employed to calculate the scattering parameters of the above network. As shown in~\Fig{circuito}, the nodes $n_{1}$ and $n_{2}$ share the same voltage, given by $(1 + R)=T$ (coming from the superposition of the incident and reflected waves at the discontinuity plane). The voltages in the rest of the lines are then
\begin{align}
  V_{00}^{\cp} &= (1 + R) = T\\
  V_{00}^{\xp} &= (1 + R) \frac{N_{00}^{\xp}}{N_{00}^{\cp}}
\end{align}
with associated currents given by
\begin{align}
  I_{00}^{\cpR} &= TY_{00}^{\cpR}\\
  I_{00}^{\xpL} &= (1 + R)\frac{N_{00}^{\xp}}{N_{00}^{\cp}}Y_{00}^{\xpL}\\
  I_{00}^{\xpR} &= T \frac{N_{00}^{\xp}}{N_{00}^{\cp}} Y_{00}^{\xpR}\;.
\end{align}

Defining now the corresponding arrays of power wave amplitudes~\cite{Pozar}, the following generalized scattering parameters are obtained:
\begin{align}
   \label{S11co} 
  S_{11}^{\cp \to \cp} &= R \\
     \label{S21co} 
  S_{21}^{\cp \to \cp} &= T\sqrt{\frac{Y_{00}^{\cpR}}{Y_{00}^{\cpL}}} \\
  \label{S11cross} 
  S_{11}^{\cp \to \xp} &=  
  (1 + R)\frac{N_{00}^{\xp}}{N_{00}^{\cp}}\sqrt{\frac{Y_{00}^{\xpL}}{Y_{00}^{\cpL}}} \\
  \label{S21cross} 
  S_{21}^{\cp \to \xp} &=  T \frac{N_{00}^{\xp}}{N_{00}^{\cp}}\sqrt{\frac{Y_{00}^{\xpR}}{Y_{00}^{\cpL}}}\;.
\end{align}

\begin{figure}
\begin{center}
\subfigure[]{
\includegraphics[width=0.45\columnwidth]{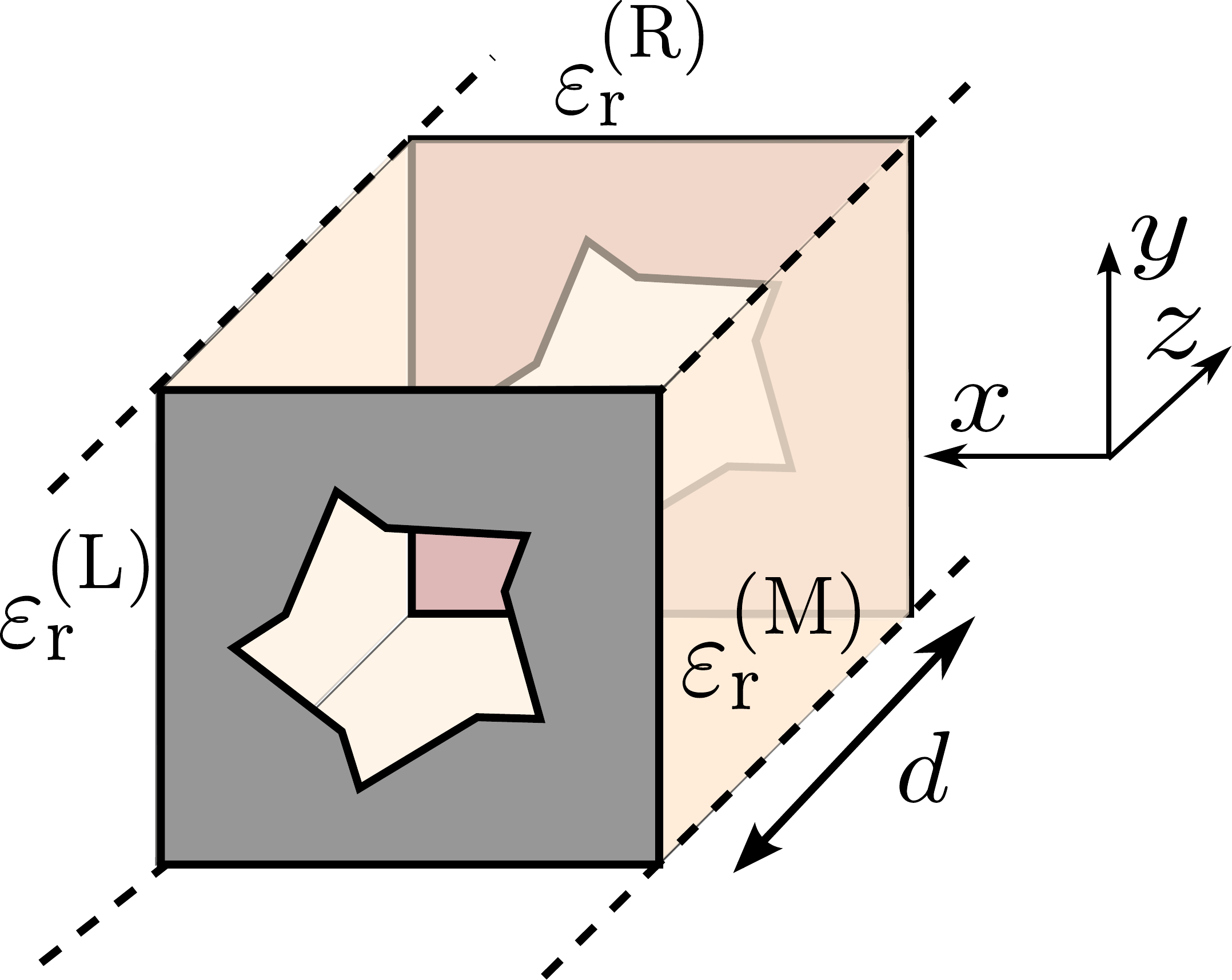}}%
\subfigure[]{
\includegraphics[width=0.45\columnwidth]{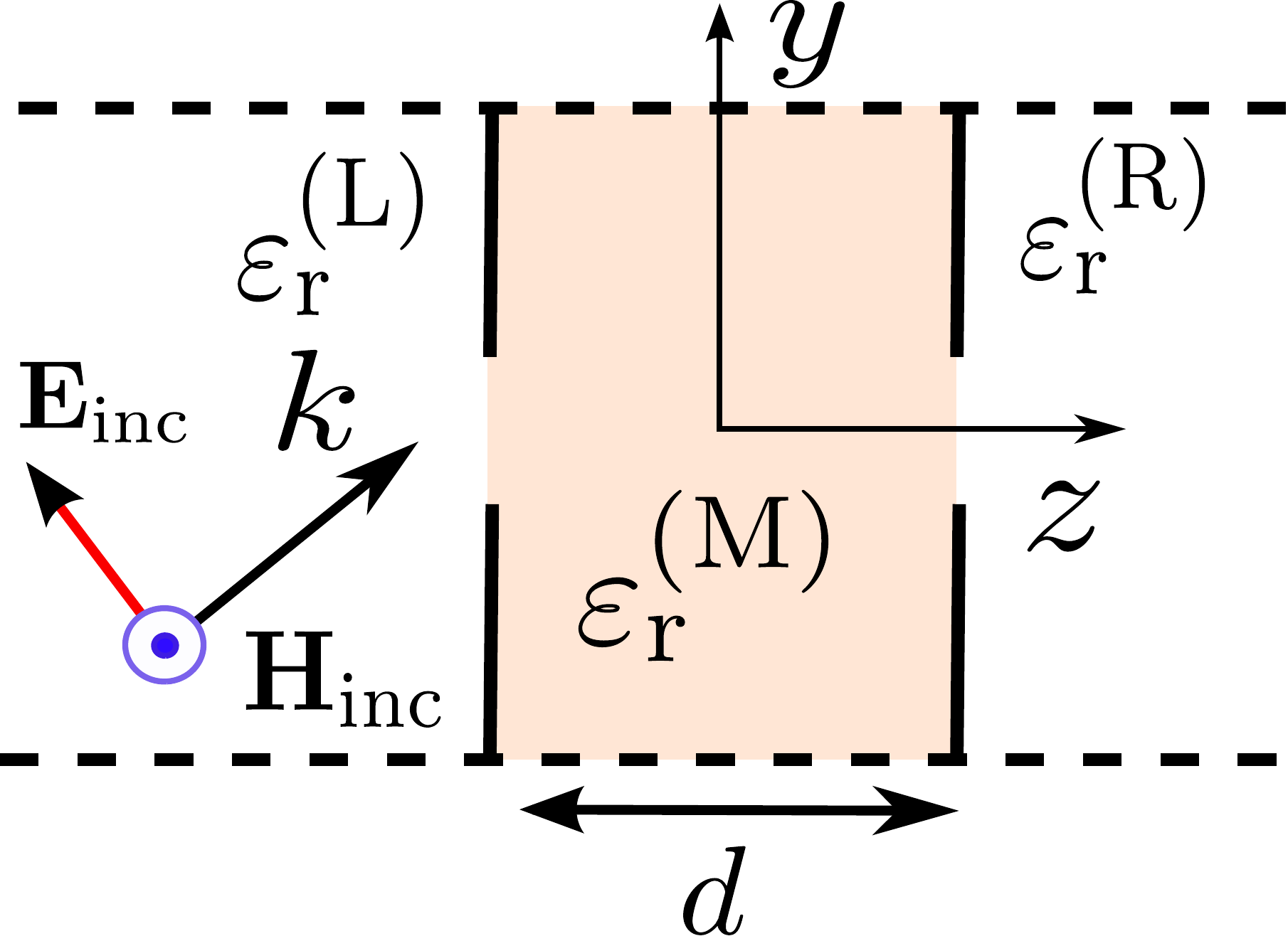}}
\end{center}
\caption{\label{double_array} (a): Unit cell in perspective. (b): Unit cell in the YZ-plane. } 
\end{figure}
\subsection{Stacks of apertures}

Stacks of two (or more) identical or nonidentical coupled apertures can also be analyzed via the equivalent circuit approach, provided that they have the same $p_x\times p_y$ dimensions (namely, the size of the unit cell should be the same). An example of this scenario is shown in~\Fig{double_array}(a). Two metallic perforated screens are separated by a dielectric material with relative permittivity $\varepsilon_{\text{r}}^{\text{(M)}}$ and thickness~$d$.  The unit cell now contains two semi-infinite dielectric slabs of relative permittivity $\varepsilon_{\text{r}}^{\text{(L/R)}}$ and the finite slab with $\varepsilon_{\text{r}}^{\text{(M)}}$, as illustrated in~\Fig{double_array}(b). 

\begin{figure}[t]
\begin{center}
\includegraphics[width=0.98\columnwidth]{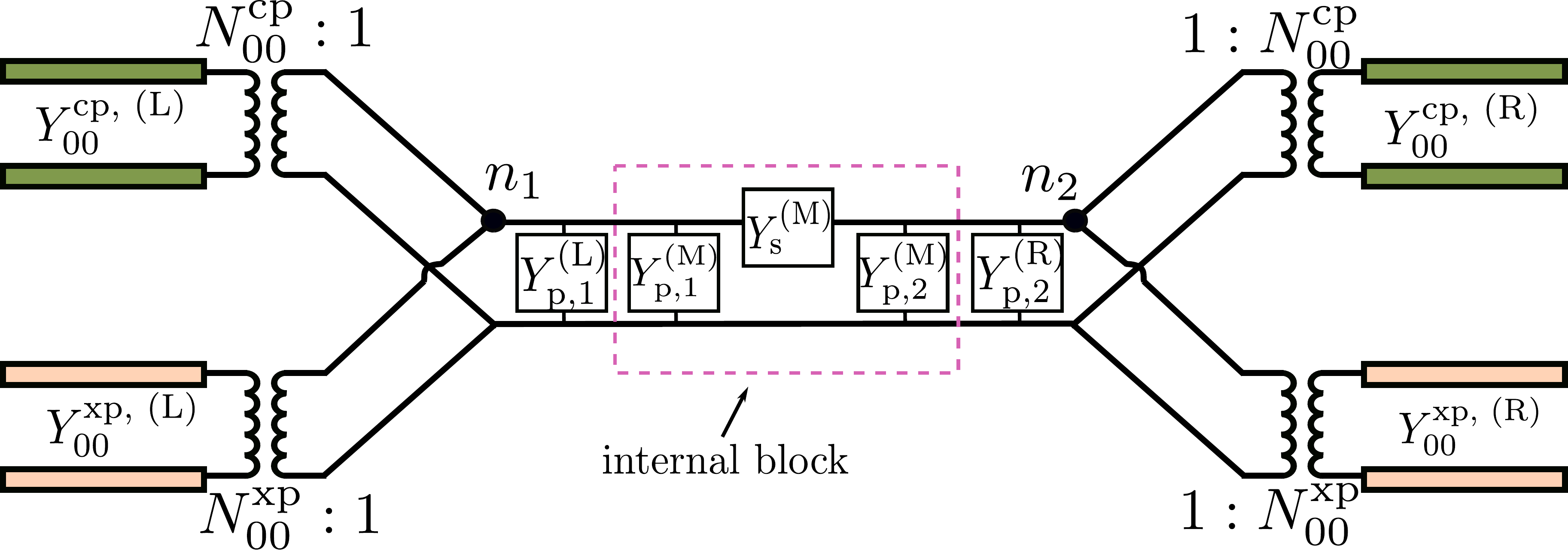} 
\end{center}
\caption{\label{circuito_double} Equivalent circuit for stacked aperture arrays.} 

\end{figure}

Following a similar rationale as in~\cite{Molero2017_asymmetrical}, the obtaining of the circuit network for this structure is based on the single-array case. Each individual aperture taking part in the stack admits to be represented by the circuit in~\Fig{circuito}. Both individual circuits can be joined by connecting the transmission lines associated with each one of the harmonics inside the finite dielectric slab (namely, connecting \textit{all} the lines associated with harmonics of the same order). 
By means of this operation and after several manipulations, the final equivalent circuit can be represented by the classical $\pi$-network shown in~\Fig{circuito_double}. Again, we have separated the co- and cross-polarized terms in the leftmost and rightmost dielectric media by its corresponding transmission lines. The admittances forming the $\pi$-block can be split into two contributions. On the one hand, the external contribution given by the harmonics $n,m\neq 0,0$ in the leftmost and rightmost dielectric media,
\begin{align}
  Y_{\text{p}, 1}^{\text{(L)}} &= \displaystyle\sump_{n, m} |N_{nm}^{\cp, (1)}|^{2} Y_{nm}^{\cpL} +\displaystyle\sump_{n, m} |N_{nm}^{\xp, (1)}|^{2} Y_{nm}^{\xpL} \\
  Y_{\text{p}, 2}^{\text{(R)}} &= \displaystyle\sump_{n, m} |N_{nm}^{\cp, (2)}|^{2} Y_{nm}^{\cpR} +\displaystyle\sump_{n, m} |N_{nm}^{\xp, (2)}|^{2} Y_{nm}^{\xpR}
\end{align}
where the transformers now include the superscript (1/2), referring to the leftmost or rightmost aperture, respectively. On the other hand, the internal block includes the information about all the possible electromagnetic couplings between the two aperture arrays. This is due to the multi-modal nature of the elements that form both the internal and external blocks. In this case, the internal block can be expressed as a $\pi$-network topology whose elements are given by
\begin{multline}
  Y_{\text{p}, 1}^{\text{(M)}} = \mathrm{j} \bigg[\displaystyle\sum_{\forall n, m} |N_{nm}^{\cp, (1)}|^{2} Y_{nm}^{\cp,\text{(M)}} +\displaystyle\sum_{\forall n, m} |N_{nm}^{\xp, (1)}|^{2} Y_{nm}^{\xp,\text{(M)}}\bigg] \\ \times \tan(\beta_{nm}^{\text{(M)}}d/2)
\end{multline}
\vspace*{-8mm}
\begin{multline}
  Y_{\text{p}, 2}^{\text{(M)}} = \mathrm{j} \bigg[\displaystyle\sum_{\forall n, m} |N_{nm}^{\cp, (2)}|^{2} Y_{nm}^{\cp,\text{(M)}} +\displaystyle\sum_{\forall n, m} |N_{nm}^{\xp, (2)}|^{2} Y_{nm}^{\xp,\text{(M)}}\bigg] \\ \times  \tan(\beta_{nm}^{\text{(M)}}d/2)
\end{multline}
\vspace*{-10mm}
\begin{multline}
   Y_{\text{s}}^{\text{(M)}} = -\mathrm{j} \bigg[\displaystyle\sum_{\forall n, m} N_{nm}^{\cp, (1)}[N_{nm}^{\cp, (2)}]^{*} Y_{nm}^{\cp,\text{(M)}}  \\ +\displaystyle\sum_{\forall n, m} N_{nm}^{\xp, (1)}[N_{nm}^{\xp, (2)}]^{*} Y_{nm}^{\xp,\text{(M)}}\bigg] \csc(\beta_{nm}^{\text{(M)}} d)\,.
\end{multline}

The topology of the circuit for the stack is slightly more complex than the circuit for the single case. The expressions for the scattering parameters are those in~\eqref{S21co}, \eqref{S11cross} and~\eqref{S21cross}. However, it can be inferred from~\Fig{circuito_double} that the nodes $n_{1}$ and $n_{2}$ do not share now the same voltage, that is, $(1 + R) \ne T$. A simple circuit analysis reveals that the voltages $(1 + R)$ and~$T$ are related by the following expression:
\begin{equation}
T=(1+R)\, \frac{Y_{\text{s}}^{\text{(M)}}}{Y_{\text{s}}^{\text{(M)}} + Y_{\text{in}}^{\text{(R)}}}
\end{equation}
with
\begin{equation}
Y_{\text{in}}^{\text{(R)}}= Y_{\text{p}, 2}^{\text{(M)}} + Y_{\text{p}, 2}^{\text{(R)}} + |N_{00}^{\cp, (2)}|^{2} Y_{00}^{\cpR} +  |N_{00}^{\xp, (2)}|^{2} Y_{00}^{\xpR}\;.
\end{equation}

This new value of $T$ can be introduced in \eqref{S21co} and~\eqref{S21cross} in order to achieve the final expressions for the generalized scattering parameters.  

\section{Periodic patch arrays. Single and stack.}
\label{Patch-arrays}

\begin{figure}
\begin{center}
\subfigure[]{
\includegraphics[width=0.45\columnwidth]{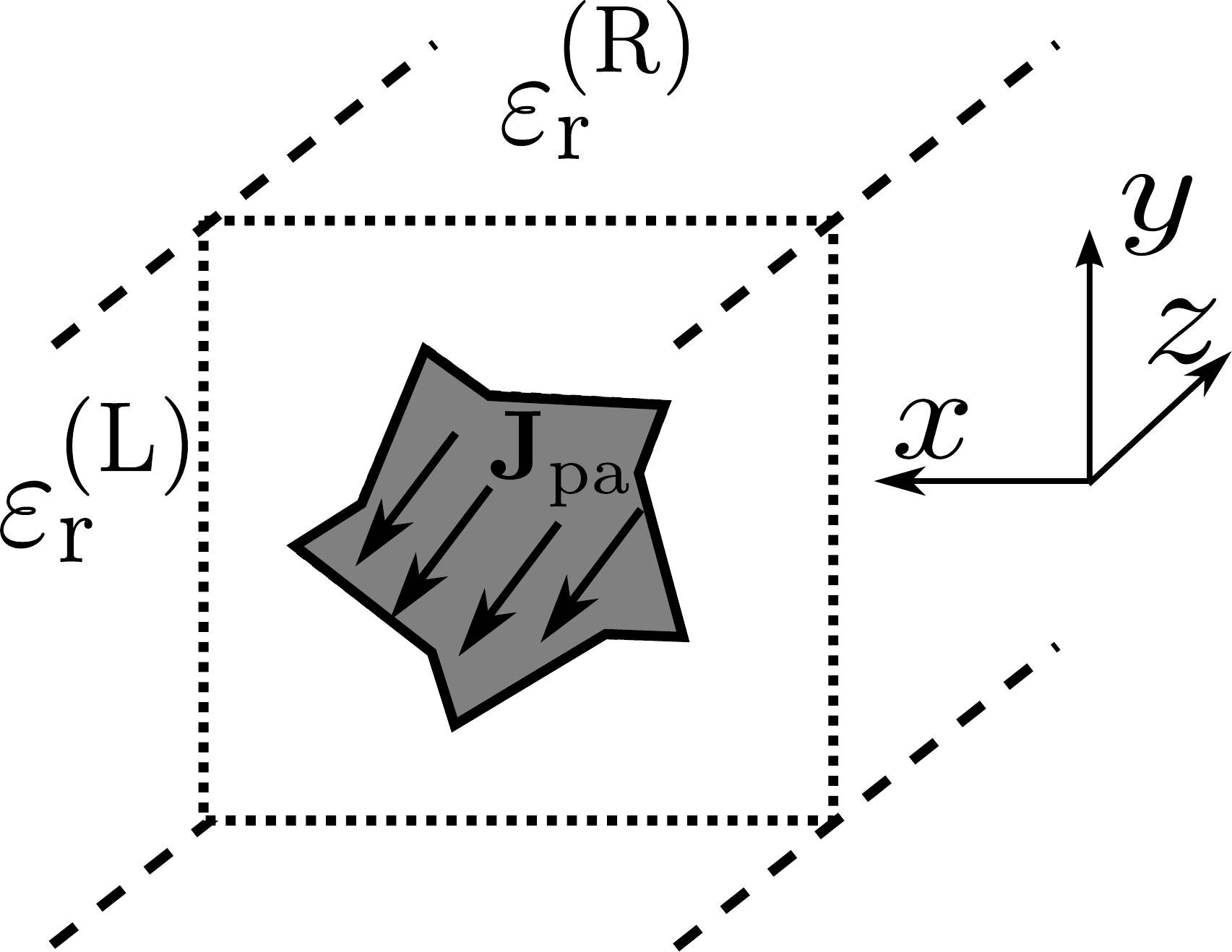}}\;%
\subfigure[]{
\includegraphics[width=0.45\columnwidth]{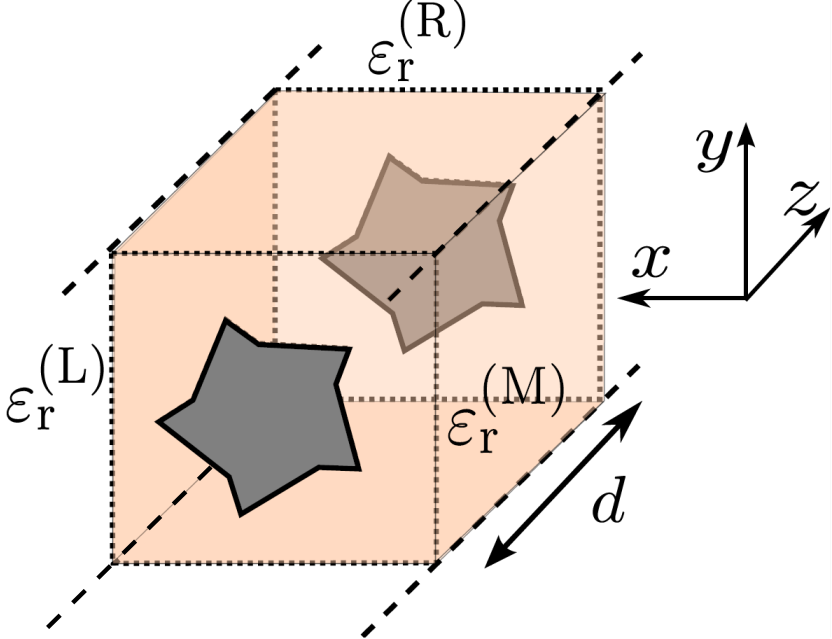}}
\end{center}
\caption{\label{patches} 
Unit cell for (a): a single-patch array, (b): a pair of coupled arrays of patches.} 
\end{figure}

\begin{figure}
\begin{center}
\subfigure[]{
\includegraphics[width=0.8\columnwidth]{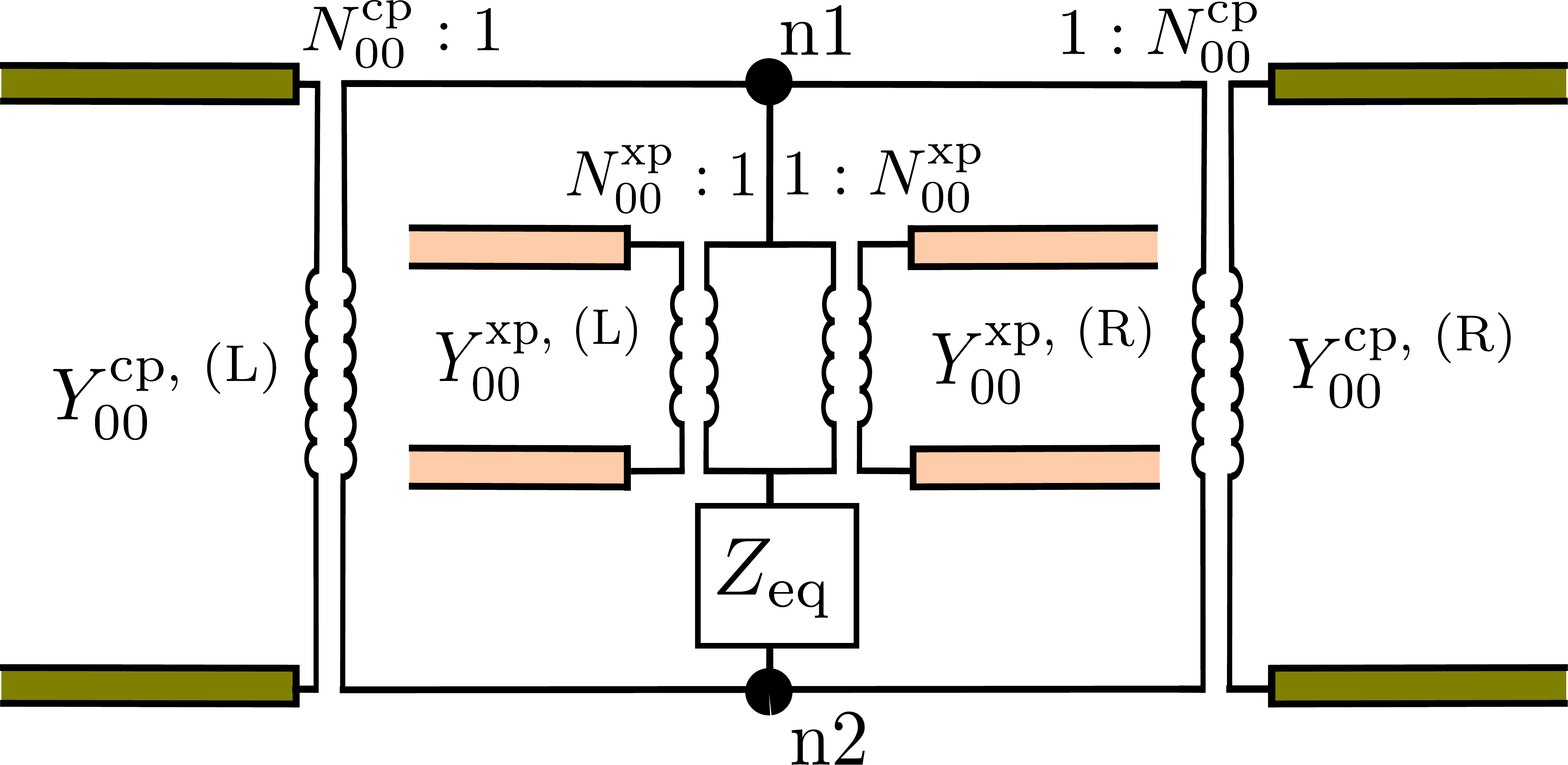}}
\subfigure[]{
\includegraphics[width=1\columnwidth]{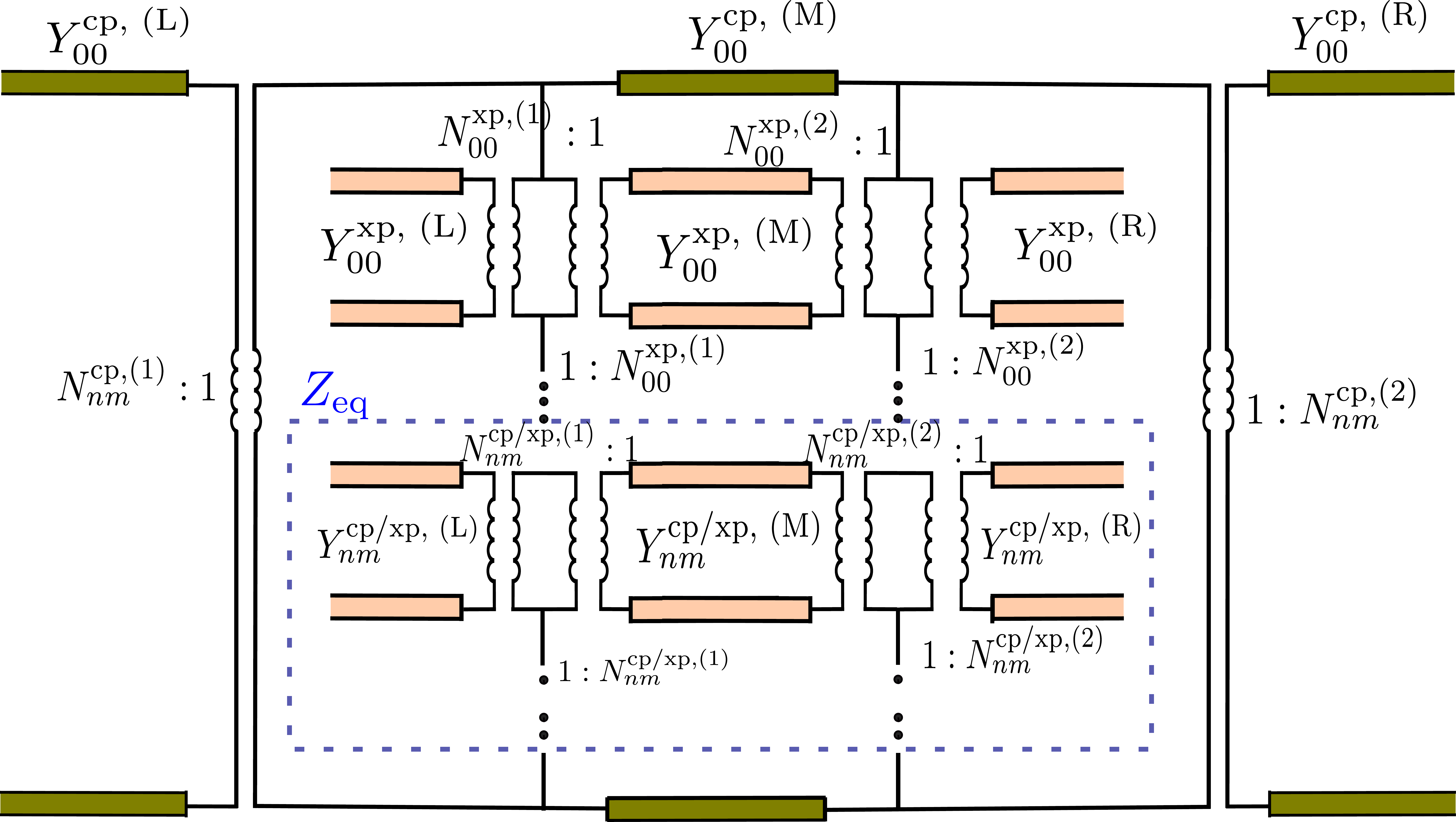}}
\end{center}
\caption{\label{circuitos_patch} 
(a): Equivalent circuit for the single array. (b): Equivalent circuit including voltages when assuming the incidence of an harmonic (labelled as the co-polarized one).} 
\end{figure}

Problems involving patches arrays can be conceived as  complementary to the aperture case and, thus, its treatment can follow the general guidelines given above. Figs.\,\ref{patches}(a) and~(b) illustrate the unit-cell problem for single and coupled patch arrays, respectively. Focusing on the single-array case, the derivation of the reflection coefficient is achieved after imposing the following two conditions at the discontinuity plane, specifically in the area of the metallic patch~$\Omega$:
\begin{gather}
    \label{cond1}  
     \hat{\mathbf{z}} \times [\mathbf{H}_t^{\text{(R)}} (x, y; \omega) - \mathbf{H}_t^{\text{(L)}}(x, y;\omega)] = \mathbf{J}_{\text{pa}}(x, y; \omega)\\
    \label{cond2}  
     \mathbf{E}_t^{(\text{L})}(x, y;\omega) = 0 = \mathbf{E}_t^{(\text{R})}(x, y;\omega) 
\end{gather}
where $\mathbf{J}_\text{pa}$ is the surface current density on the patch.

The transverse magnetic and electric fields are described in terms of Floquet harmonics as in~\eqref{eq:EL}, \eqref{eq:ER}, \eqref{H1} and \eqref{H2}. The surface current is assumed to be given by a spatial distribution independent from frequency, namely,
\begin{equation}\label{eq:Jpa}
\mathbf{J}_{\text{pa}}(x,y;\omega) = B(\omega)\Jpa(x, y)\,.  
\end{equation}
After manipulating \eqref{cond1} and \eqref{cond2}, taking into account~\eqref{eq:Jpa}, the following expression relating the amplitudes of the harmonics and the reflection and transmission coefficients of the incident one is found: 
\begin{equation}\label{VnmcocrP}
    V_{nm}^{\text{cp/xp}}\frac{Y_{nm}^{\text{cp/xp, (L)}} + Y_{nm}^{\text{cp/xp, (R)}} }{N_{nm}^{\cp/\xp}} =  \frac{- (1 - R) Y_{00}^{\cpL} + TY_{00}^{\cpR}}{N_{00}^{\cp}}
\end{equation}
where $T = (1 + R)$,
and
\begin{equation}\label{FourierJ}
    N_{nm}^{\text{cp/xp}} = \widetilde{\Jpa}(k_{xn}, k_{ym})\cdot \hat{\mathbf{e}}_{nm}^{\text{cp/xp}}
\end{equation}
with $\widetilde{\Jpa}(k_{xn}, k_{ym})$ being the 2-D Fourier transform of $\Jpa(x,y)$ at $(k_{xn}, k_{ym})$. The reflection coefficient $R$ can be written as
\begin{equation}\label{Rpatch}
R = \frac{\dfrac{Y_{00}^{\cpL} - Y_{00}^{\cpR}}{|N_{00}^{\text{cp}}|^{2}} - \bigg[\dfrac{|N_{00}^{\xp}|^{2}}{Y_{00}^{\xpL} + Y_{00}^{\xpR} }  + Z_{\text{eq}}\bigg]^{-1}}{\dfrac{Y_{00}^{\cpL} + Y_{00}^{\cpR}}{|N_{00}^{\text{cp}}|^{2}} + \bigg[\dfrac{|N_{00}^{\xp}|^{2}}{Y_{00}^{\xpL} + Y_{00}^{\xpR} }  + Z_{\text{eq}}\bigg]^{-1}}      
\end{equation}
with
\begin{equation}\label{eq:Zeq}
    Z_{\text{eq}} = \displaystyle\sump_{n, m}  \dfrac{|N_{nm}^{\cp}|^{2}}{Y_{nm}^{\cpL} + Y_{nm}^{\cpR} } + \displaystyle\sump_{n, m}  \dfrac{|N_{nm}^{\xp}|^{2}}{Y_{nm}^{\xpL} + Y_{nm}^{\xpR} }\;.
\end{equation}
As in the aperture-array case, $Z_{\text{eq}}$ includes all the harmonics of orders $n, m$ except  those with $n, m = 0,0$. Similarly, electromagnetic coupling between adjacent cells is taken into consideration by the multi-modal nature of the circuit in ~\Fig{circuitos_patch}(a). However, at the light of~\eqref{eq:Zeq}, the connection among all these elements is now in \textit{series} in contrast to the parallel connections observed in the aperture case. The cross-polarization elements of order $n,m = 0,0$ also take part in the group of elements connected in series. 


Following the theory in~\cite{Pozar} and using the expression for $V_{nm}^{\cp/xp}$ obtained after manipulating \eqref{VnmcocrP}, the generalized scattering parameters for the case of a single patch-based FSS are found to be
\begin{align}
   \label{S11coPB} 
  S_{11}^{\cp \to \cp} &= R \\
     \label{S21coPB} 
  S_{21}^{\cp \to \cp} &= (1 + R)\sqrt{\frac{Y_{00}^{\text{cp,(R)}}}{Y_{00}^{\text{cp,(L)}}}} \\
  \label{S11crossPB} 
  S_{11}^{\cp \to \xp} &=  \frac{N_{00}^{\xp}}{N_{00}^{\cp}} \frac{ (R-1) Y_{00}^{\cpL} + TY_{00}^{\cpR}}{Y_{00}^{\text{xp, (L)}} + Y_{00}^{\text{xp, (R)}}} \sqrt{\frac{Y_{00}^{\text{xp,(L)}}}{Y_{00}^{\text{cp,(L)}}}}
   \\
  \label{S21crossPB} 
  S_{21}^{\cp \to \xp} &=  \frac{N_{00}^{\xp}}{N_{00}^{\cp}} \frac{(R-1) Y_{00}^{\cpL} + TY_{00}^{\cpR}}{Y_{00}^{\text{xp, (L)}} + Y_{00}^{\text{xp, (R)}}} \sqrt{\frac{Y_{00}^{\text{xp,(R)}}}{Y_{00}^{\text{cp,(L)}}}}
\end{align}


The couple of patch-arrays separated by a distance $d$ depicted in~\Fig{patches}(b) also admits to be represented by an equivalent circuit. Similarly as in the aperture-case, we proceed by connecting transmission lines associated with individual harmonics. \Fig{circuitos_patch}(b) shows the final topology. The transmission lines related to the cross-pol term of order $n = m = 0$ in inner medium (M) are connected. This procedure is also realized by the rest of higher order harmonics (all of them included in $Z_{\text{eq}}$ in~\Fig{circuitos_patch}(a)). The resulting topology is not appropriate to construct individual blocks and derive a final $\pi$-network. Furthermore, the complexity of calculating the scattering parameters for the cross-pol term  increases considerably. A possible way to address this calculation appeals to transfer-matrix notation. In particular, the composition of scattering (S) matrices  has been the method employed. Each S-matrix describes and represents the coupling between all the harmonics excited in an individual patch array. The final S-matrix of the system is the result of the composition of all the individual matrices. The transmission/reflection coefficients of the cp- and xp-components are directly extracted from this final scattering matrix.
It is worth remarking that the calculation is still cumbersome and no physical insight is inferred from the process.

\section{Numerical examples}
\label{Numerical examples}

In this section, some numerical examples of FSS structures are given in order to validate the analytical circuit approach and show its accuracy and computational efficiency. We compare the results with commercial full-wave simulators as well as with previously published works in the literature. Examples of aperture and patch arrays (single and stacked layers) are reported.

\subsection{Aperture-based Circular Polarizer}

\begin{figure}
\begin{center}
\subfigure[]{
\hspace*{0.5cm}
\includegraphics[width=0.7\columnwidth]{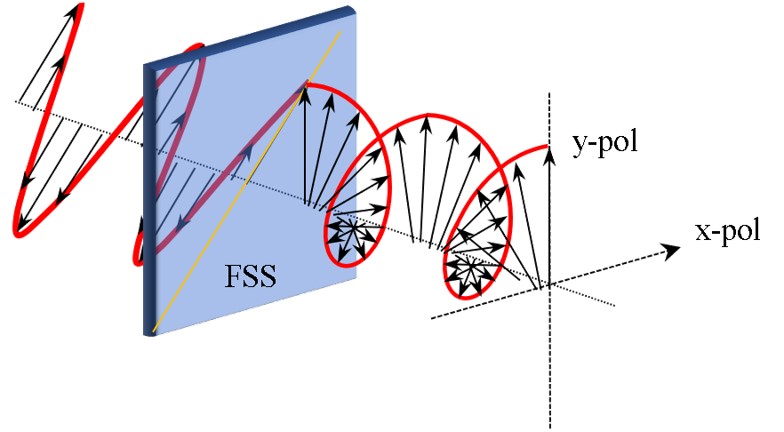}}
\subfigure[]{
\includegraphics[width=0.65\columnwidth]{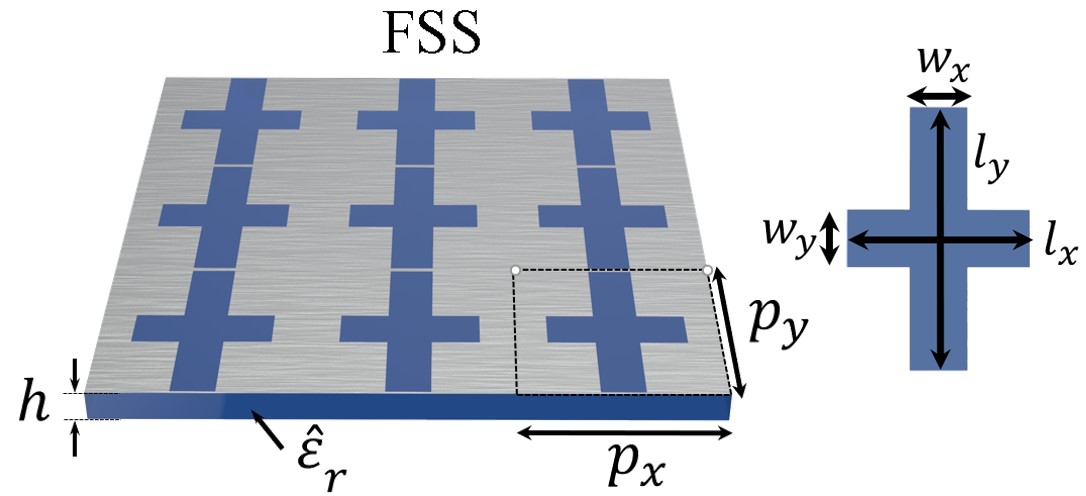}}
\hspace*{-0.2cm}
\subfigure[]{
\includegraphics[width=0.3\columnwidth]{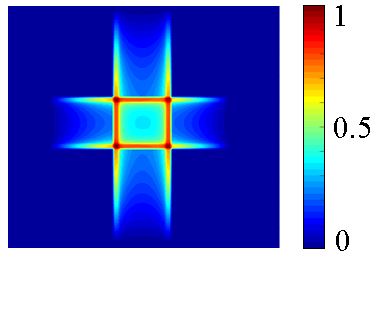}}
\subfigure[]{
\includegraphics[width=1.02\columnwidth]{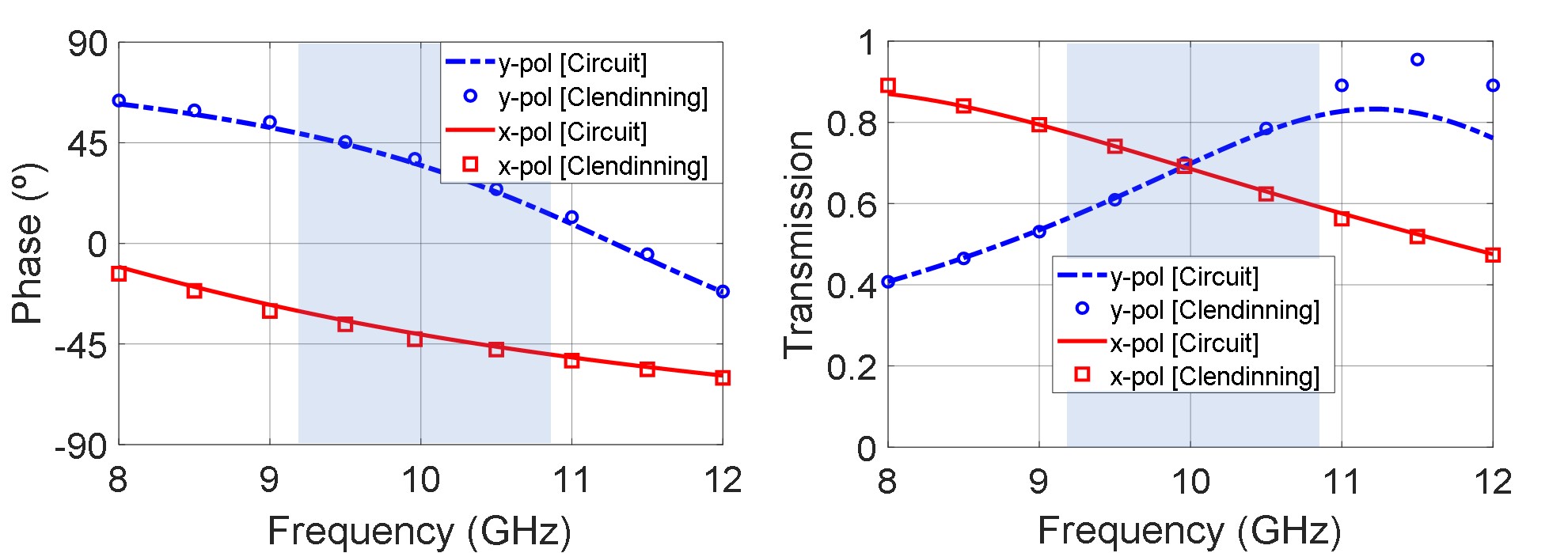}}
\end{center}
\caption{\label{CP_cross} (a) Linear-to-circular polarization converter.  (b) Single-layer aperture-based FSS presented in \cite{clendinning} acting as a circular polarizer. (c) Spatial profile considered at the aperture, computed with \eqref{eq:profile_cross}. (d)~Phase and amplitude of the transmission parameters ($x$-pol and $y$-pol waves). The shadowed area indicates the frequency range where the transmitted wave is circularly polarized (AR$\leq3$ dB).  Losses are considered in the dielectric, $\hat{\varepsilon}_r = {\varepsilon}_r(1-\jj\tan \delta)$, $\varepsilon_r=2.95$,  $\tan\delta =0.025$.} 
\end{figure}

The first structure under analysis is the aperture-based linear-to-circular polarizer presented in \cite{clendinning}. The operation of the structure as well as its forming layers can be visualized in \Figs{CP_cross}(a) and (b), respectively. The FSS is formed by a periodic array of cross-shaped apertures lying on a lossy dielectric substrate. When a $45\degree$ linearl$y$-polarized plane wave impinges the structure, the transmitted $x$-pol and $y$-pol components are delayed by $\pi/2$ radians while maintaining the same level of amplitude. This is due to the different lengths that the two magnetic dipoles forming the cross have, which leads to a circularl$y$-polarized wave at the output.

The analytical spatial profile of the electric field assumed in the cross-shaped aperture, displayed in \Fig{CP_cross}(c), is mathematically modeled as \cite{basis_functions}
\begin{equation} \label{eq:profile_cross}
  \boldsymbol{\mathcal{E}}(x,y)=\dfrac{\cos\!\left(\pi y/l_y\right)}{\sqrt{1-\left(2x/w_x\right)^2}}\,   \hat{\mathbf{x}}\, 
  +\, \dfrac{\cos\!\left(\pi x/l_x\right)}{\sqrt{1-\left(2y/w_y\right)^2}} \,  \hat{\mathbf{y}}\;.
\end{equation}
In the above expression, the horizontal ($y$-oriented $E$-field) and vertical ($x$-oriented $E$-field) magnetic dipoles are independently modeled. 

The phase and amplitude coefficients of the transmitted $x$-pol and $y$-pol waves are plotted in~\Fig{CP_cross}(d) and compared with the measurements shown in \cite{clendinning}. As observed, there is a good agreement between the ECA results and the experimental data. For the computation of 501 frequency points, the circuit approach took only 2.28 seconds. This demonstrates the efficiency of the present model. Furthermore, it should be remarked the ability of the equivalent circuit to easily account for dielectric losses in the substrate. Slight differences are seen in the amplitude coefficient at the end of the displayed frequency range. This is attributed to the large size of the cross-shaped apertures as well as the field coupling between horizontal and vertical dipoles, not considered in expression \eqref{eq:profile_cross}. The shadowed area in \Fig{CP_cross}(d) highlights the frequency range where the axial ratio is below~3\,dB; namely, where the transmitted wave is circularly polarized from a practical point of view. 

\subsection{Asymmetric Loaded Magnetic Dipoles}
\begin{figure}
\begin{center}
\subfigure[]{
\includegraphics[width=0.37\columnwidth]{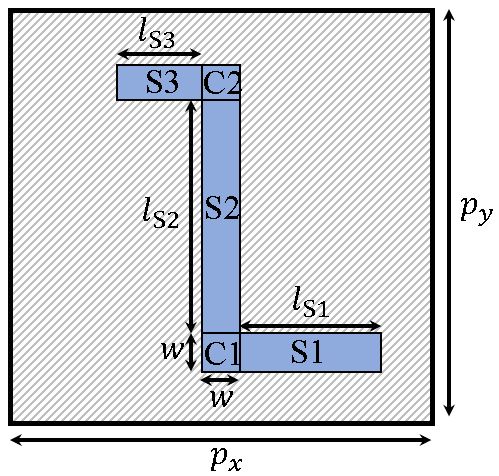}}
\hspace*{4mm}
\subfigure[]{
\includegraphics[width=0.46\columnwidth]{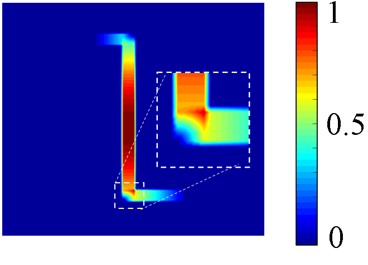}}
\subfigure[]{
\includegraphics[width=0.98\columnwidth]{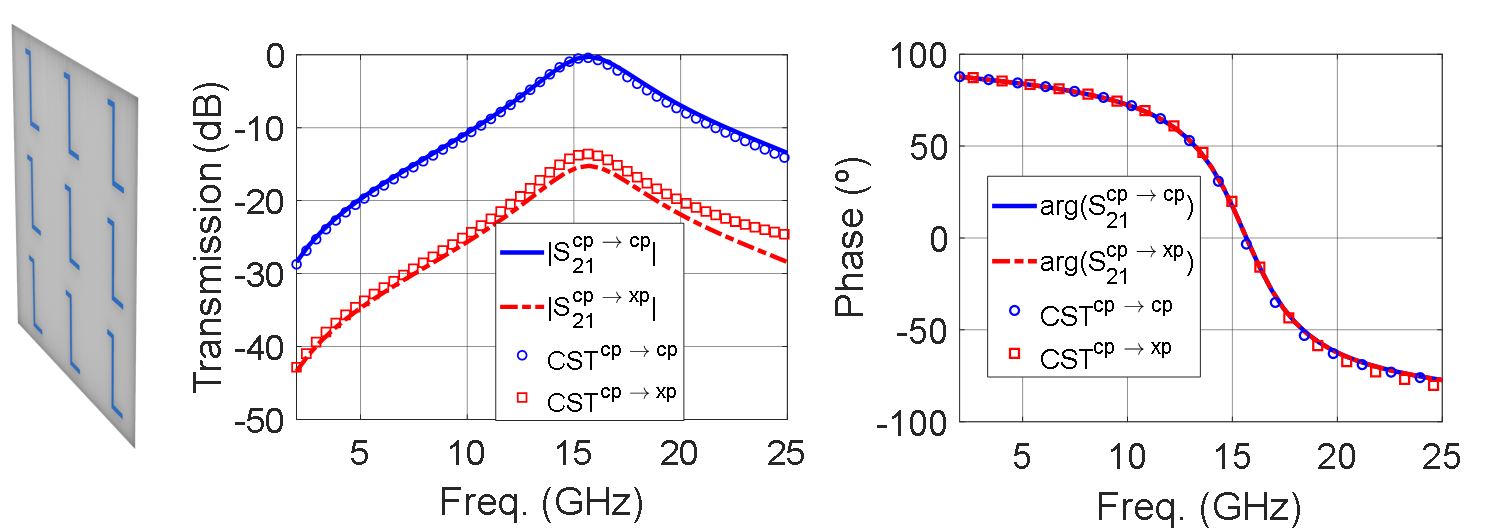}}
\subfigure[]{
\includegraphics[width=0.98\columnwidth]{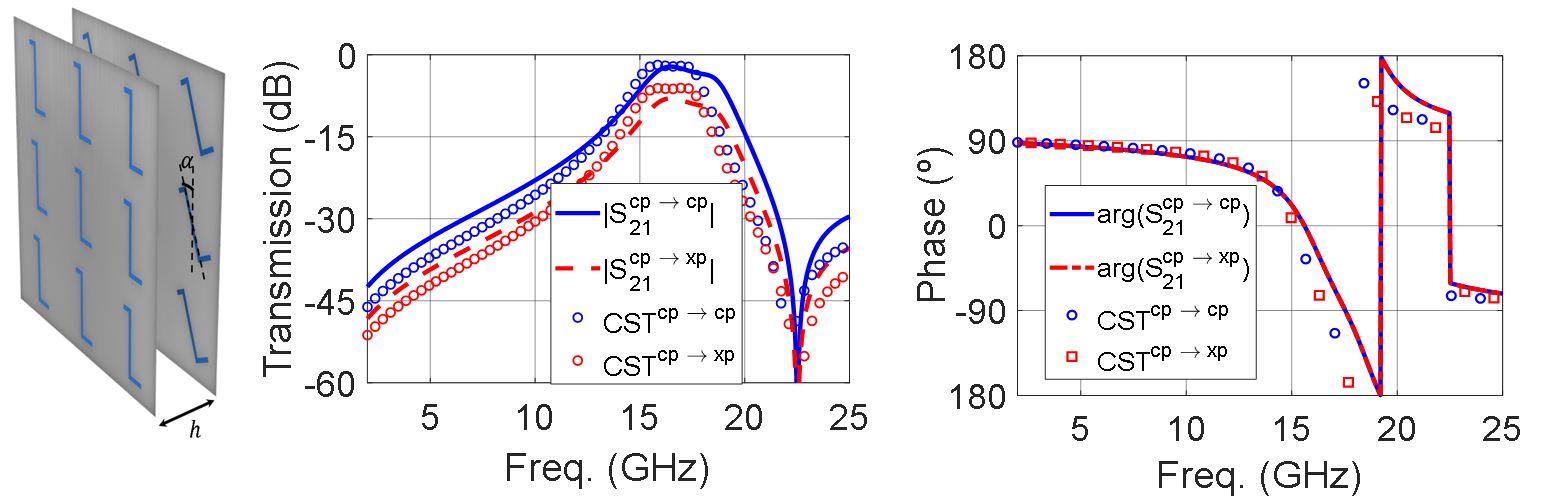}}
\end{center}
\caption{\label{fig:Stack1}  (a) Front view of the unit cell containing the asymmetrical loaded magnetic dipoles. (b) Spatial profile considered at the aperture, computed with \eqref{eq:profile_dipole}, and a detail of a corner section.  (c) Single array. (d) Two-layer stack whose plates are rotated 20 degree. TE normal incidence is assumed. Geometrical parameters: $p_x=p_y=10$\,mm, $l_{S1}=2$\,mm, $l_{S2}=6$\,mm, $l_{S3}=1$\,mm,  $w=0.5$\,mm, $\alpha=20\degree$, and $h=2.5$\,mm.}  
\end{figure}

The second example is inspired by the synthesis method of anisotropic FSS reported in~\cite{Costa-AP2020}, based on the use of equivalent circuits. The FSS structure is formed by asymmetric loaded magnetic dipoles, as can be seen in~\Fig{fig:Stack1}(a). As discussed in~\cite{meander} for meander-line gratings, a frequency-independent sine-shaped (half sine) spatial profile can be assumed to be excited at the aperture. The considered profile is constant along the slot width~$w$ and is mathematically modeled as
\begin{equation} \label{eq:profile_dipole}
\boldsymbol{\mathcal{E}}(x,y)= \boldsymbol{\mathcal{E}_\text{sec}}(x,y) + \boldsymbol{\mathcal{E}_\text{cor}}(x,y)
\end{equation}
where
\begin{equation}
\boldsymbol{\mathcal{E}_\text{sec}}(x,y)= 
\begin{cases}
\sin(\frac{\pi x}{L})\,  \hat{\mathbf{y}}\;,  \quad &\text{in region S1} \\[5pt]
\sin(\frac{\pi [y+L_1]}{L})\,  \hat{\mathbf{x}}\; ,  \quad &\text{in region S2}\\[5pt] 
\sin(\frac{\pi [x+L_2]}{L})\, \hat{\mathbf{y}}\;, \quad &\text{in region S3} \\
\end{cases} 
\end{equation}
takes into account the sinusoidal variation along the different longitudinal sections [see~\Fig{fig:Stack1}(a)] of the loaded dipole ($L_1=l_{S1}+w$, $L_2=l_{S1}+2w+l_{S2}$), and
\begin{equation}
\boldsymbol{\mathcal{E}_\text{cor}}(x,y)=
\begin{cases}
 \boldsymbol{\mathcal{E}_\text{cor}}^\text{(C1)}(x,y)\;, \quad \text{in corner C1}\\
 \boldsymbol{\mathcal{E}_\text{cor}}^\text{(C2)}(x,y)\;, \quad \text{in corner C2}\\
\end{cases}
\end{equation}
includes the boundary conditions at the corner sections, with
\begin{multline} \label{eq_corner1}
\boldsymbol{\mathcal{E}_\text{cor}}^\text{(C1)}(x,y) = 
\left(1-\frac{y}{w}\right) \sin\!\left(\frac{\pi L_1}{L}\right)\hat{\mathbf{x}}  \\  
+ \left(1-\frac{x}{w}\right) \sin\!\left(\frac{\pi l_{S1}}{L}\right)\hat{\mathbf{y}}
\end{multline}
\vspace*{-4mm}
\begin{multline}
\boldsymbol{\mathcal{E}_\text{cor}}^\text{(C2)}(x,y)  = 
\left(1-\frac{y}{w}\right) \sin\!\left(\frac{\pi L'_2}{L}\right)\hat{\mathbf{x}}  \\  
+ \left(1-\frac{x}{w}\right) \sin\!\left(\frac{\pi L_2}{L}\right)\hat{\mathbf{y}}
\label{eq_corner2}
\end{multline}
and $L'_2=l_{S1}+w+l_{S2}$. Note that the total length of the loaded dipole is $L=l_{S1}+l_{S2}+l_{S3}+2w$. The magnitude of the spatial profile can be observed in \Fig{fig:Stack1}(b). As appreciated in \eqref{eq_corner1},\eqref{eq_corner2}, a linear decay that maintains the continuity of the fields has been assumed at the corner sections. The same consideration is valid when dealing with currents in patch structures, as in Fig.~\ref{fig:doubleL}. Although this approach may seem simple, it has proven to give accurate results \cite{meander}. 

\Fig{fig:Stack1}(c) illustrates the transmission parameters (amplitude and phase) of the co-pol (cp$\rightarrow$cp) and cross-pol (cp$\rightarrow$xp)  terms for a single, free-standing layer when a TE-polarized plane wave illuminates the FSS. An excellent agreement is observed between our ECA results and the commercial software~CST.  It should be noted that the length of the horizontal arms of the dipoles control the cross-pol level; namely, the larger the horizontal arms are, the higher the cross-pol term is. 

A variant of the above structure is formed when the previous layer is stacked together with a second rotated layer  of equal dimensions (angle of rotation $\alpha=20^\degree$). In our equivalent circuit approach, the incorporation of this second layer to account for the spatial profile of the rotated aperture simply demands a linear transformation of the original spatial profile~\eqref{eq:profile_dipole}, as pointed out in~\cite{multimodalTAP}.  The transmission parameters (amplitude and phase) of the stack are plotted in~\Fig{fig:Stack1}(d), where a good agreement of our ECA results is observed with CST data, despite the complexity of the FSS under analysis. Using the same computer, the circuit model took~2.02 and 3.18~seconds to compute the scattering parameters at 501~frequency points of the single array  and the two-layer stack, respectively, while CST (default mesh size, 60~Floquet harmonics) took~6 and 8~minutes in these calculations. It is then apparent that, within its validity range, the ECA is a very efficient procedure to study the scattering properties of FSSs as well as an excellent tool for optimization tasks where many computations are required.

\subsection{Double L-shaped dipole}

\begin{figure}[t]
\begin{center}
\subfigure[]{\includegraphics[width=0.95\columnwidth]{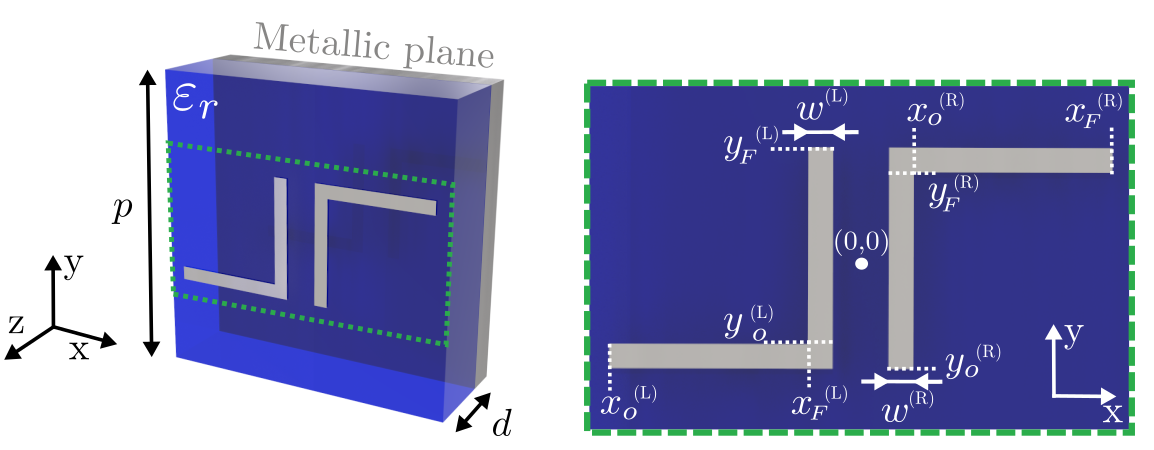}}
\subfigure[]{\includegraphics[width=0.9\columnwidth]{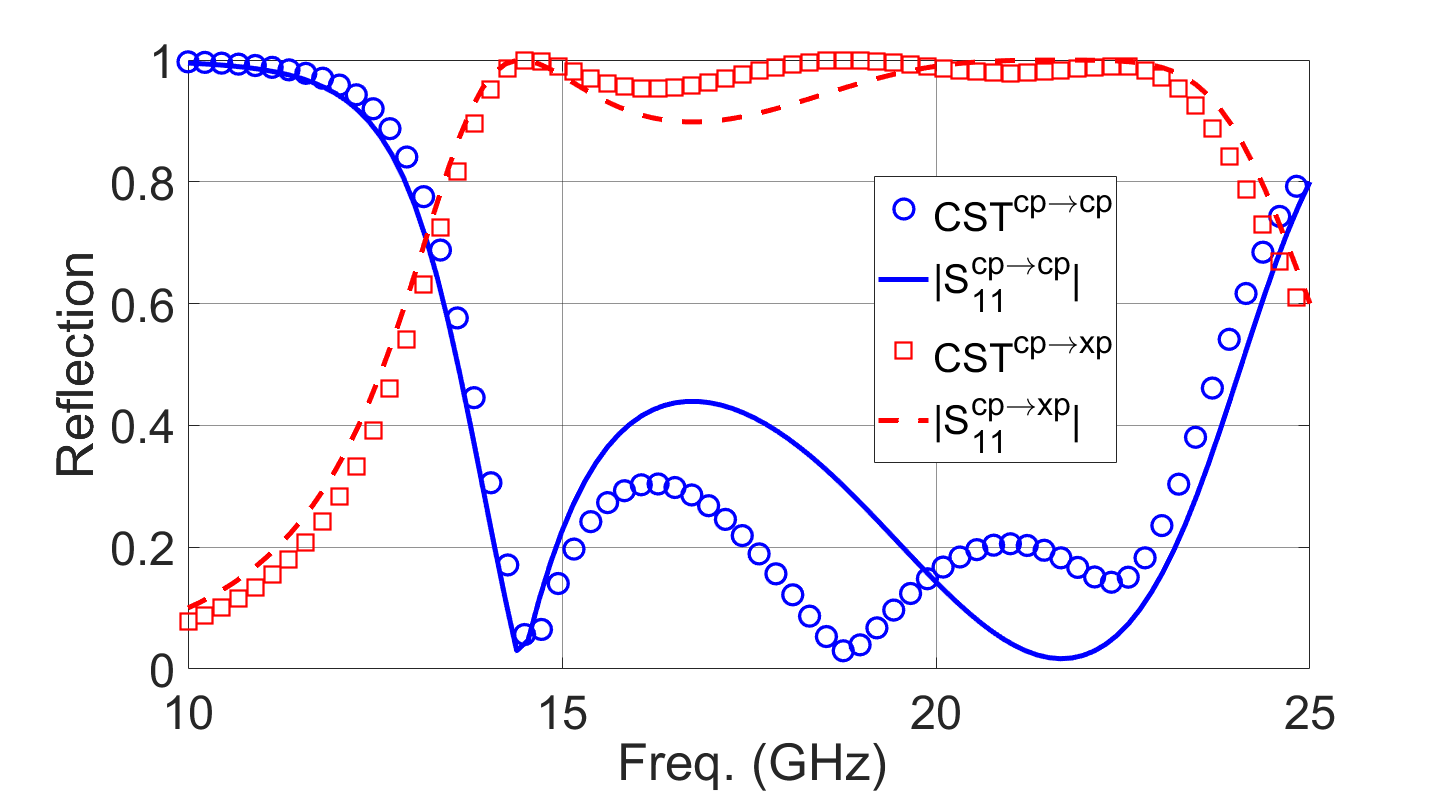}}
\end{center}
\caption{\label{fig:doubleL}  (a): Unit cell in perspective and under front view.  (b):~$|S_{11}|$ data obtained by the ECA and CST for TE normal incidence. Structure parameters (lengths in mm): $p = 4.5$, $d = 1.2$, $w^{\text{(L)}}~=~w^{(\text{R})}~=~0.2$, $x_{0}^{(\text{L})} = -2.225$, $x_{\text{F}}^{\text{(L)}} = -0.625$, $y_{0}^{\text{(L)}} = -0.7$, $y_{\text{F}}^{\text{(L)}} = 0.9$, $x_{0}^{(\text{R})} = 0.625$, $x_{\text{F}}^{\text{(R)}} = 2.225$, $y_{0}^{\text{(R)}} = -0.9$, $y_{\text{F}}^{\text{(R)}} = 0.7$, $\varepsilon_{\text{r}} = 10$.   } 
\end{figure}

A first example of a patch-array periodic structure to be analyzed with the circuit model is inspired in the structure reported in~\cite{DobleL}. The unit cell, shown in~\Fig{fig:doubleL}(a), comprises two identical L-shaped metallic dipoles printed on a grounded dielectric slab. Each individual dipole will be labelled as $\text{(L)}$ or $\text{(R)}$, referring to the \emph{left}- or \emph{right}-side  position in the cell. The (L)-dipole is  rotated $180\degree$ with respect to the (R)-dipole, thus creating an asymmetric unit-cell pattern. The presence of the ground plane ensures full-reflection operation.

In the frame of the circuit model approach described in Sec.\,\ref{Patch-arrays}, the ground plane is included in the circuit in~\Fig{circuitos_patch} just by short-circuiting the lines associated with the rightmost medium with permittivity~$\varepsilon_{\text{r}}^{(\text{R})}$. The length of all these short-circuited transmission lines coincides with the dielectric-slab thickness. As required by the ECA, the spatial profile of the surface current  distribution, $\Jpa(x, y)$, in each of the L-shaped dipoles is assumed to  be frequency independent. The longitudinal component of the current is taken to have the form of a half-sine function, distributed along the corresponding strip, and constant along the strip width. Likewise the example in the previous section, the surface-current function is mathematically expressed in two parts:
\begin{equation}
 \Jpa^{(\text{L/R})}(x, y) = \Jpa_{\text{sec}}^{(\text{L/R})}(x, y) + \Jpa_{\text{cor}}^{(\text{L/R})}(x, y)  
\end{equation}
where 
\begin{equation}
    \Jpa^{(\text{L})}_{\text{sec}}(x, y) = 
    \begin{cases}
    \sin\!\left(\dfrac{\pi x}{L^{\text{(L)}}}\right)  \hat{\mathbf{x}}\;,  \quad &\text{in  S1} \\[5pt]
    \sin\!\left(\dfrac{\pi (d_{\text{x}}^{\text{(L)}} + w^{\text{(L)}} + y - y^{\text{(L)}}_{0})}{L^{\text{(L)}}}\right)  \hat{\mathbf{y}}\;,  \quad &\text{in S2}
    \end{cases}
\end{equation}
\begin{equation}
    \Jpa_{\text{sec}}^{(\text{R})}(x, y) =  
    \begin{cases}
    \sin\!\left(\dfrac{\pi y}{L^{\text{(R)}}}\right)  \hat{\mathbf{y}}\;,  \quad &\text{in S1} \\[5pt]
    \sin\!\left(\dfrac{\pi (d_{\text{y}}^{\text{(R)}} + w^{\text{(R)}} + x - x^{\text{(R)}}_{0})}{L^{\text{(R)}}}\right)  \hat{\mathbf{x}}\;,  \quad &\text{in S2}
\end{cases}
\end{equation}
and 
\begin{multline}
 \Jpa^{\text{(L)}}_{\text{cor}}(x, y) = \sin\!\left(\dfrac{\pi d_{\text{x}}^{\text{(L)}}}{L^{\text{(L)}}}\right) \dfrac{x - x_{\text{F}}^{\text{(L)}}}{w^{\text{(L)}}}\, \mathbf{\hat{x}} \\ 
 + \sin\!\left(\dfrac{\pi (d_{\text{x}}^{\text{(L)}} + w^{\text{(L)}})}{L^{\text{(L)}}}\right) \dfrac{y - (y_{\text{F}}^{\text{(L)}} - w^{\text{(L)}})}{w^{\text{(L)}}}\, \mathbf{\hat{y}} 
 \end{multline}
 \begin{multline}
 \Jpa^{\text{(R)}}_{\text{cor}}(x, y) = \sin\!\left(\dfrac{\pi d_{\text{y}}^{\text{(R)}}}{L^{\text{(R)}}}\right) \dfrac{y - y_{\text{F}}^{\text{(R)}}}{w^{\text{(R)}}} \, \mathbf{\hat{y}} \\ 
 + \sin\!\left(\dfrac{\pi (d_{\text{y}}^{\text{(R)}} + w^{\text{(R)}})}{L^{\text{(R)}}}\right) \dfrac{x - (x_{\text{F}}^{\text{(R)}} - w^{\text{(R)}})}{w^{\text{(R)}}}\, \mathbf{\hat{x}}
 \end{multline}
 with 
 \begin{align*}
 d_{\text{x}}^{\text{(L/R)}} &= x_{\text{F}}^{\text{(L/R)}} - x_{0}^{\text{(L/R)}} \\
 d_{\text{y}}^{\text{(L/R)}} &= y_{\text{F}}^{\text{(L/R)}} - y_{0}^{\text{(L/R)}} \\
  L^{\text{(L/R)}} &= d_{\text{x}}^{\text{(L/R)}} + d_{\text{y}}^{\text{(L/R)}} + w^{\text{(L/R)}} \,.
\end{align*}
The corresponding transformer ratios $N_{nm}^{\text{cp/xp}}$ are given in terms of the 2-D Fourier Transform of $\Jpa^{\text{(L/R)}}$; namely, according to~\eqref{FourierJ} 
\begin{equation}
    N_{nm}^{\text{cp/xp}} = \left[\widetilde{\Jpa}^{\text{(L)}}(k_{xn}, k_{ym}) + \widetilde{\Jpa}^{\text{(R)}}(k_{xn}, k_{ym})\right] \cdot \hat{\mathbf{e}}_{nm}^{\text{cp/xp}}\,.
\end{equation}

It is worth remarking that if the dipoles are very close each other, some degree of mutual coupling is expected, affecting the excitation of \emph{individual} surface-current distribution in each dipole. Therefore, the individual currents in the left-and right patches are not expected to be identical to the ones of each patch acting separately. Since the proposed current profiles  assume no mutual coupling, the accuracy of the surface-current profiles  deteriorates as both dipoles get closer.     

In order to validate the circuit-model approach as well as the surface-current functions, \Fig{fig:doubleL} reproduces one of the results reported in \cite[Fig.\,7]{DobleL} where a wideband polarizer is presented. Both the left and right dipoles are identical (dimension in the caption of~\Fig{fig:doubleL}). The dielectric substrate is $1.2\,$mm thick with relative permittivity $\varepsilon_{\text{r}}^{(\text{R})} = 10$. The structure is excited by TE normal incidence (electric field directed along the $y$-axis). The reflection of this field component (cp) and the orthogonal field component (xp) are both plotted in \Fig{fig:doubleL}(c). The cell configuration is such that wideband \emph{field rotation} is achieved, as shown by the high reflection magnitude of the xp-component from 14.5\,GHz to 23\,GHz in contrast with the small reflection magnitude of the cp-component. 
The agreement between results provided by CST and those given by the circuit model is quite acceptable, with some slight differences caused by the approximated nature of the surface-current functions employed for this example. Possible sources of inaccuracies can be associated with the modelling of the current in corners, realized by including a simple linear decay in the sine function. The neglecting of  mutual-coupling effects in the surface current functions is also expected to yield some inaccuracies. In turn, the computation time required by the ECA is very short in comparison to the CPU time of CST. The ECA employed $1.13$ seconds to reproduce the solution, whereas CST needed $5$ minutes approximately. Hence, ECA stands out as a very quick and efficient analysis tool if a small degree of inaccuracy is acceptable. 

\subsection{Split-ring dipole}

\begin{figure}
\begin{center}
\subfigure[]{\includegraphics[width=0.49\columnwidth]{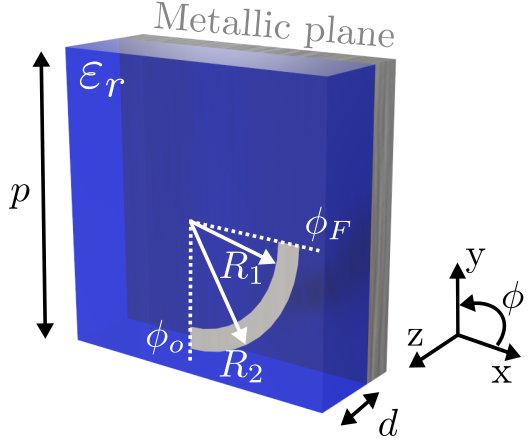}}
\subfigure[]{\includegraphics[width=0.48\columnwidth]{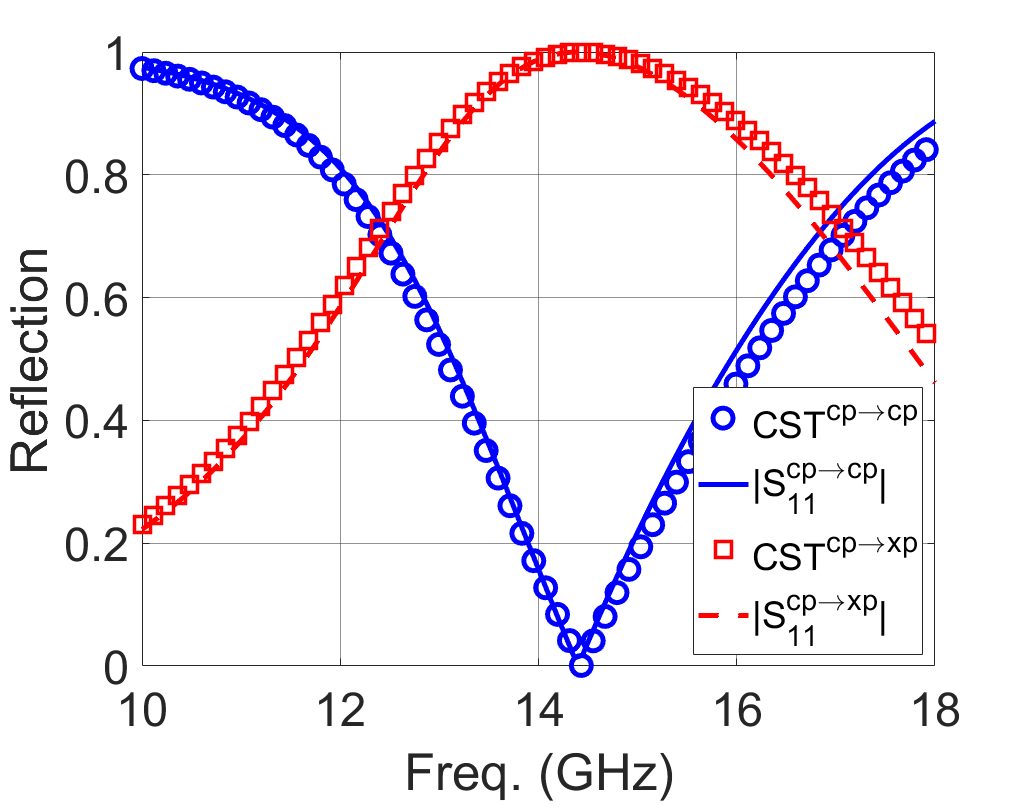}}
\subfigure[]{\includegraphics[width=0.49\columnwidth]{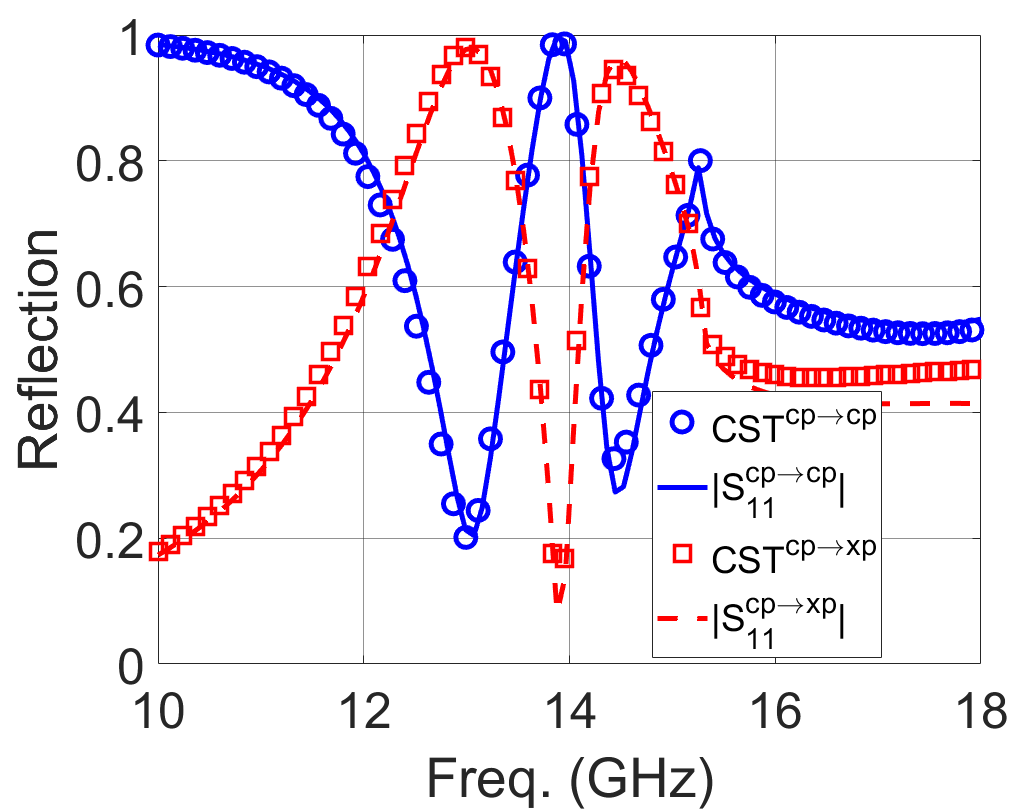}}
\subfigure[]{\includegraphics[width=0.49\columnwidth]{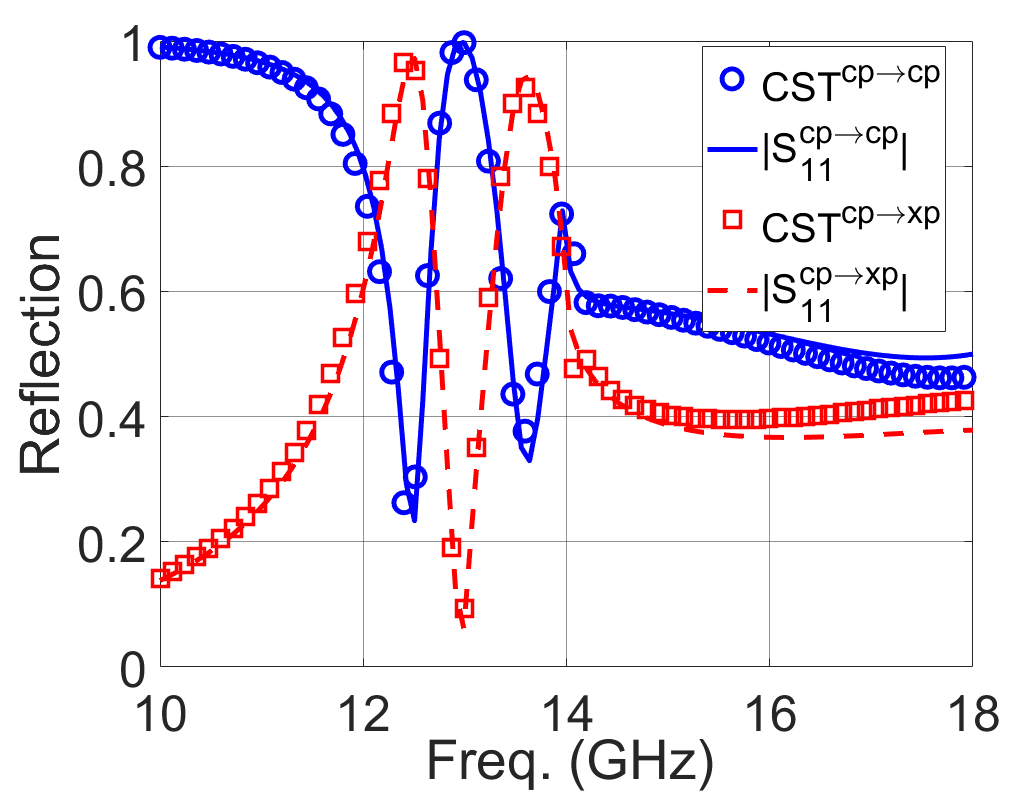}}
\subfigure[]{\includegraphics[width=0.49\columnwidth]{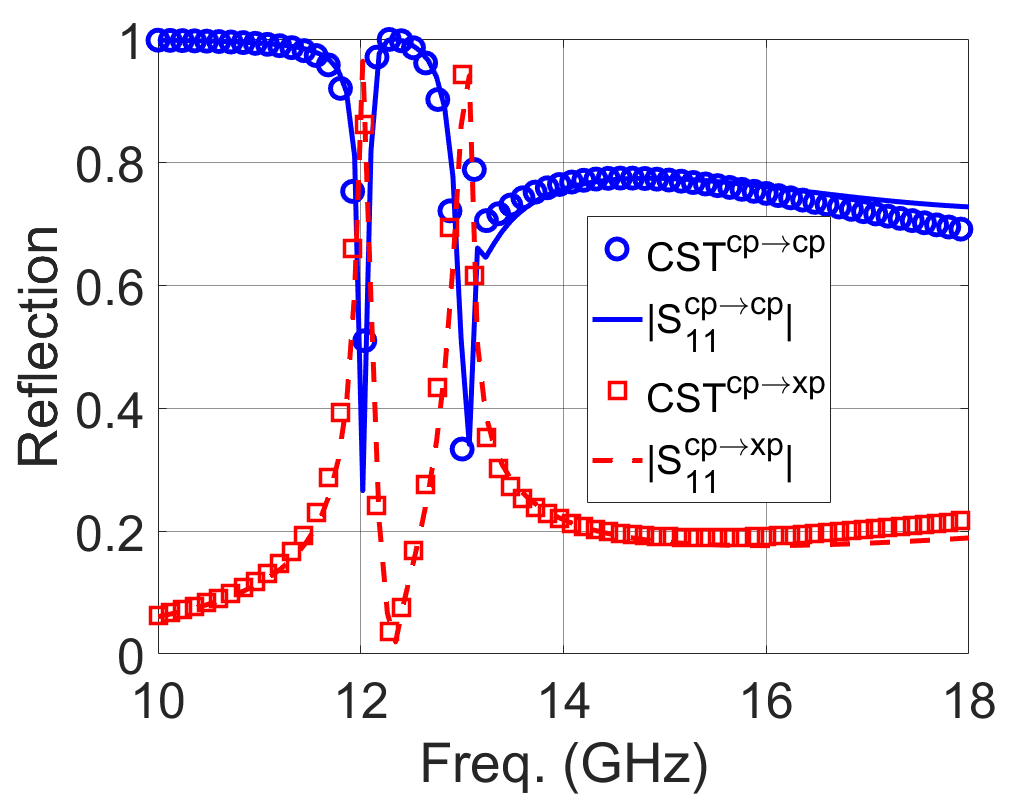}}
\end{center}
\caption{\label{fig:split_ring} TM normal and oblique incidence. (a) Unit cell. (b): Scattering parameters obtained for $\theta = 0\degree$. (c): $\theta = 45\degree$. (d): $\theta = 60\degree$. (e): $\theta = 80\degree$. Structure parameters: $p = 11.5\,$mm, $d = 3\,$mm, $R_{1} = 3.9\,$mm, $R_{2} = 4.75\,$mm, $\phi_{0} = 270\degree$, $\phi_{\text{F}} = 360\degree$, $\varepsilon_{\text{r}} = 2.55$.} 
\end{figure}

A second case study of the circuit model for patch arrays is the structure whose unit cell is shown in~\Fig{fig:split_ring}(a)-(b). The patch is  a circular ring section, defined by the arc comprised between the angles~$\phi_{0}$ and~$\phi_{\text{F}}$. The width of the patch is defined between the inner and outer radii $R_{1}$ and $R_{2}$. As the ring section is printed on a grounded dielectric slab, the corresponding equivalent circuit is topologically identical to the one employed in the previous example. The surface current is now assumed to be directed along the ring section; namely, the longitudinal component of the current takes the direction of the polar unit vector~$\boldsymbol{\hat{\phi}}$ (the current is null at the ends of the patch). A half-since function, adapted to the geometry of the patch, is again a very good option to describe the current behaviour:   
\begin{equation}
 \Jpa(R, \phi) = \sin\!\left(\pi \dfrac{\phi - \phi_{0}}{\phi_{\text{F}} - \phi_{0}}\right)  \boldsymbol{\hat{\phi}}
\end{equation}
with $\boldsymbol{\hat{\phi}} = -\sin\phi\, \mathbf{\hat{x}} + \cos\phi\, \mathbf{\hat{y}}$\,. 
For simplicity, the current is taken as constant in the transverse section comprised between $R_{1}$ and $R_{2}$. 

Analytical results obtained by the ECA and the software CST are plotted in~\Fig{fig:split_ring}(c)-(f). The magnitude of the co- and xp-component is plotted versus frequency for different incident angles~$\theta$. TM~incidence is here considered, that is, the electric-field vector lies on the $(x,z)$-plane (along the $x$-direction for $\theta= 0$). The behavior of the xp-component varies with frequency and angle. Full power transfer from the incident to the xp-component is achieved at frequencies around 14\,GHz for normal incidence. By increasing the incident angle, the spectrum begins to interchange full-reflection peaks related to the co- and xp-components. This is clearly manifested for  $45\degree, 60\degree, 80\degree$. Since the periodicity of the cell is $p = 11.5\,$mm, the onset frequency of grating lobes of first-orders ($n,m = 1,0$, $nm = 0,1$) is $f = 26.08\,$GHz. This onset frequency goes down to $15.27\,$GHz,  $13.98\,$GHz and $13.14\,$GHz for $\theta = 45\degree, 60\degree$ and $80\degree$, respectively. The agreement between CST and the ECA is quite good up to $18\,$GHz, which is beyond the onset frequencies for the largest angles. This validates the ECA as a suitable tool to work in complex scenarios, such as oblique-incidence in the diffraction zone.  

\subsection{Stack of split ring dipoles}
\begin{figure}
\begin{center}
\subfigure[]{\includegraphics[width=0.8\columnwidth]{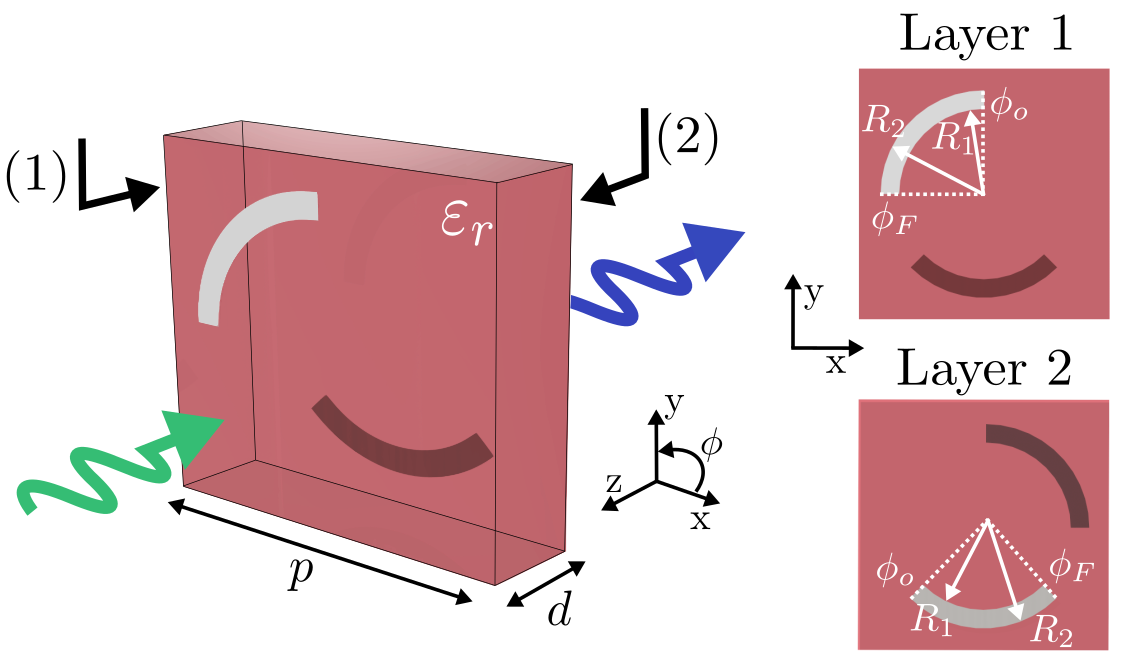}}
\subfigure[]{\includegraphics[width=0.48\columnwidth]{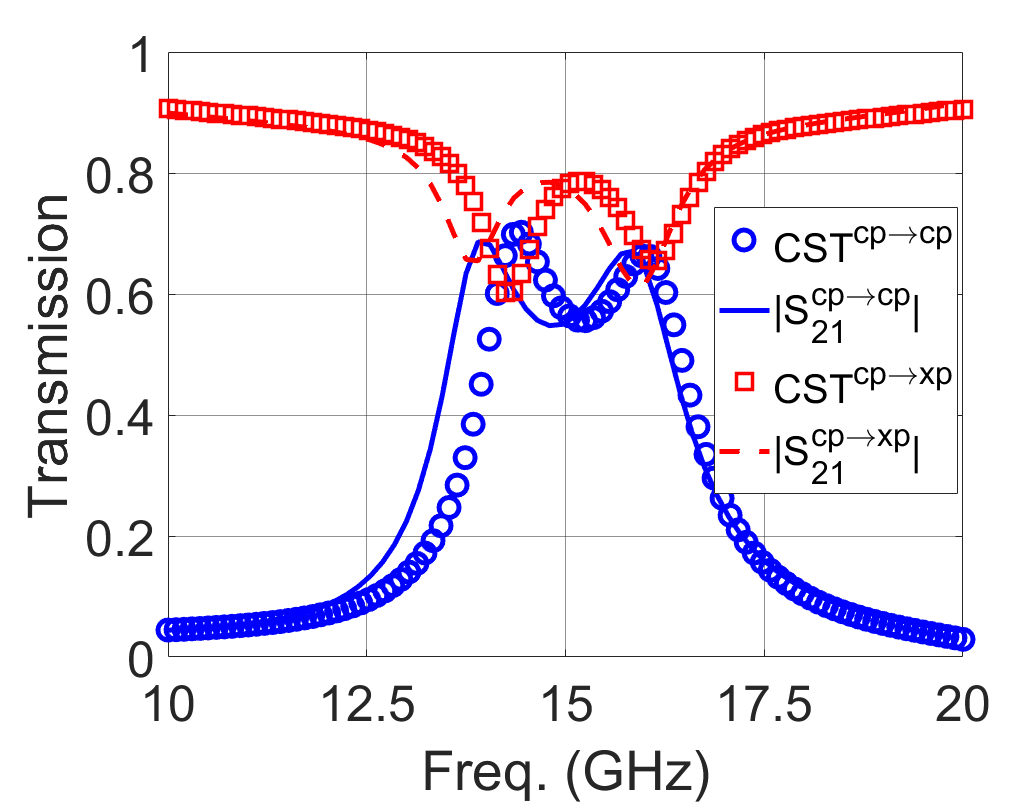}}
\subfigure[]{
\includegraphics[width=0.475\columnwidth]{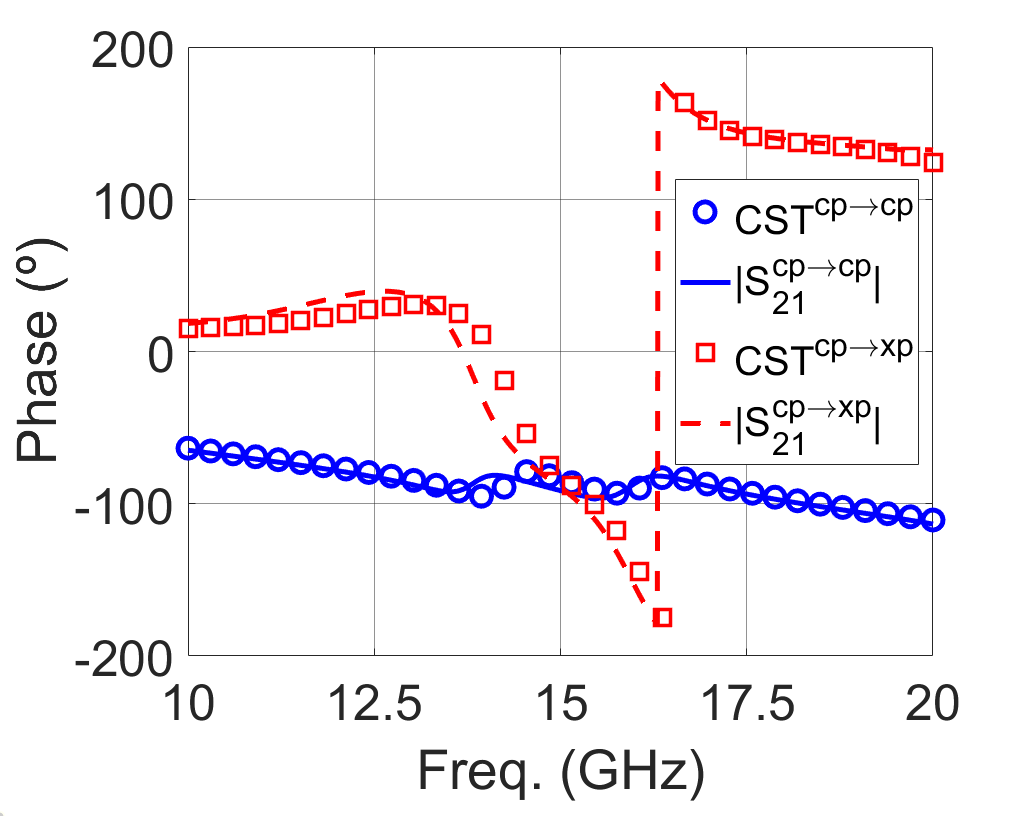}}
\subfigure[]{\includegraphics[width=0.48\columnwidth]{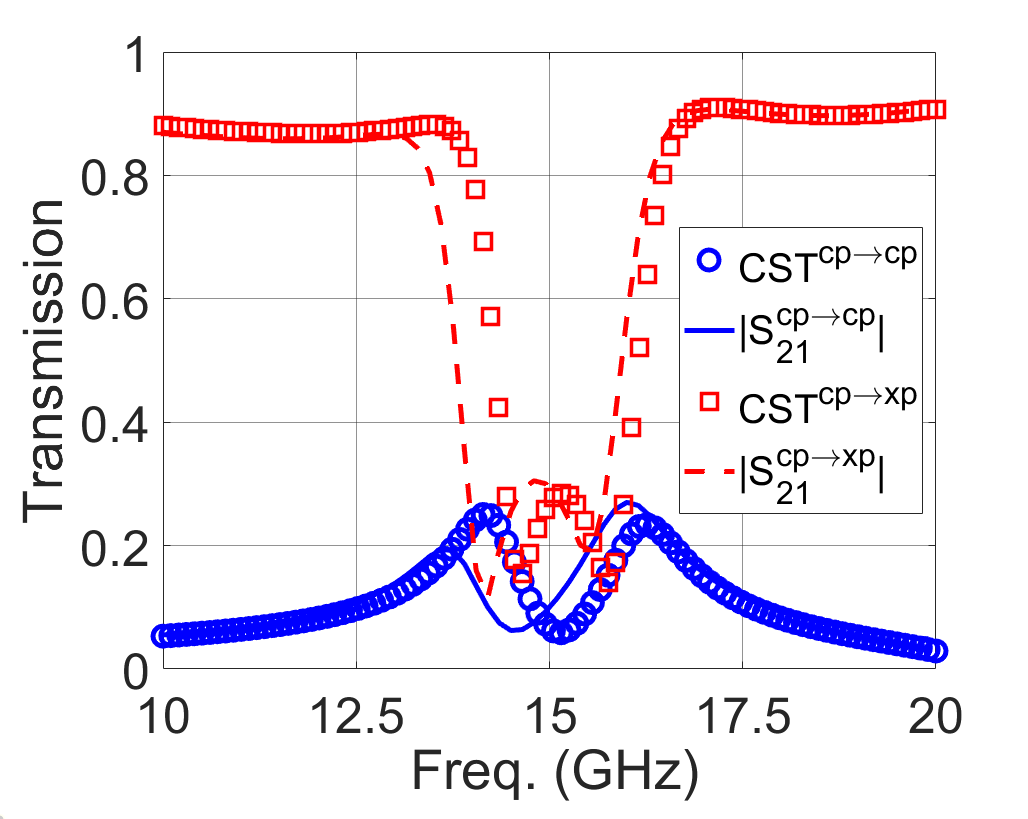}}
\subfigure[]{
\includegraphics[width=0.475\columnwidth]{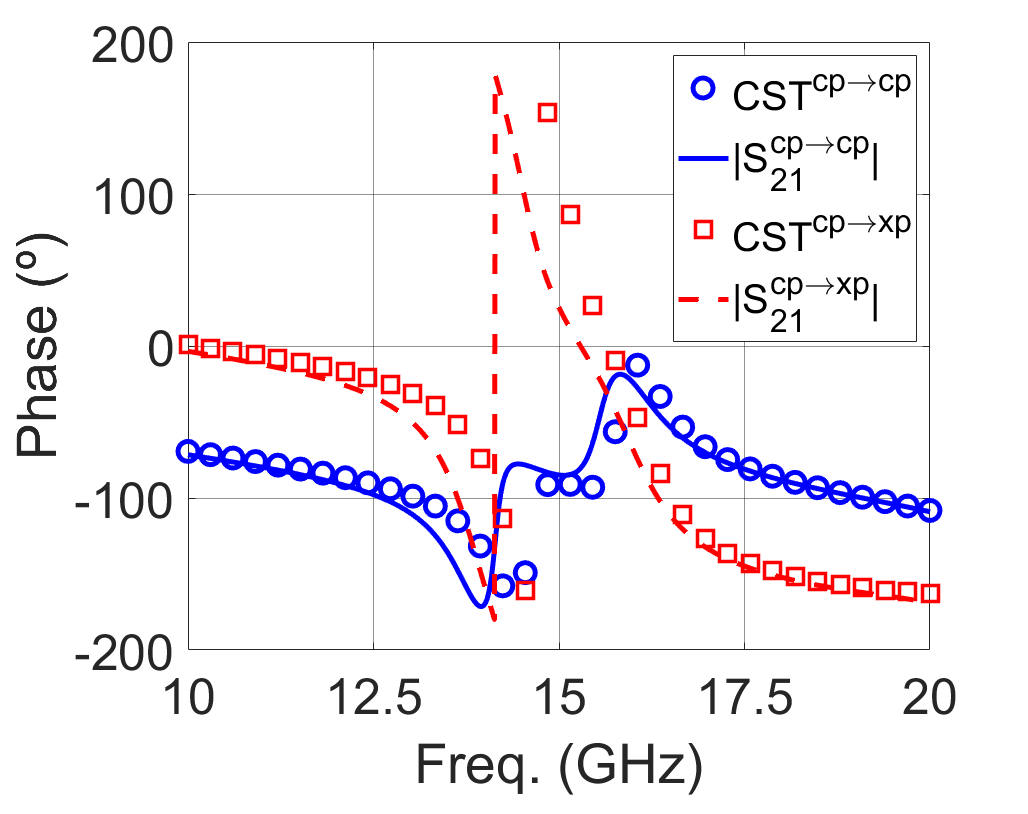}}
\end{center}
\caption{\label{fig:split_ring_stack} (a) Unit cell of the stack. The position of the ring sections in the front and back layers is fixed according to the structure parameters defined in \Fig{fig:split_ring}(a). (b)-(c)~TE and (d)-(e)~TM incidence. Common structure parameters: $p = 11.5\,$mm, $d = 3\,$mm, $\varepsilon_{\text{r}} = 2.55$, $R_{1} = 3.9\,$mm, $R_{2} = 4.75\,$mm. Patch on layer~(1): $\phi_{0} = 90\degree$, $\phi_{\text{F}} = 180\degree$. Patch on layer~(2): $\phi_{0} = 225\degree$, $\phi_{\text{F}} = 315\degree$.} 
\end{figure}

A last example of a stack of a couple of patch-based arrays is considered in~\Fig{fig:split_ring_stack}. This example is based on the ring-section geometry used in the previous section, but replacing the ground plane by a second array. \Figs{fig:split_ring_stack}(a) shows the complete unit cell, which consists of a stack of two metallic ring sections printed on opposite faces of a dielectric slab with permittivity~$\varepsilon_{\text{r}}$ and thickness~$d$. The position and length of each ring section is fixed by the initial and final angle $\phi_{0}$ and $\phi_{\text{F}}$, respectively. The width is again defined in terms of the inner and outer radii ($R_{1}$ and $R_{2}$).

\Figs{fig:split_ring_stack}(b)-(e) shows the frequency behavior of the transmission parameter (amplitude and phase) for both the co- and xp-components for TE and TM normal incidence, respectively. The unit cell has been conceived to be highly asymmetric (that is, $\phi_{0}$ and $\phi_{\text{F}}$ of the left and right ring sections do not coincide) to force the appearance of cross polarization. However its amplitude strongly depends on the incident polarization. TE~incidence, with the electric field along the~$y$-direction, tends to equalize the amplitudes of the cp- and xp-components from~14 to~16\,GHz in transmission. TM~incidence does not achieve such an increase of the xp-amplitude, and most of power that is lost by the cp-component is not transferred to the xp-component but is reflected back (results of $|S_{11}|$ are not plotted in the figure). The agreement between the results obtained by CST and the ECA is quite good in both cases. It is worth remarking the high complexity of the corresponding equivalent circuit for this scenario, as can be appreciated in \Fig{circuitos_patch}(b). The CPU time needed for the ECA was about $3$ seconds, contrasting with CST which needed around 8 minutes to provide a solution.  
 
\section{Conclusion}
\label{sec:Conc}
In this paper, we have derived an analytical equivalent circuit approach, based on a Floquet modal expansion, for the characterization of the co-pol and cross-pol scattering components in FSS structures formed by periodic arrays of patches and apertures in single and stacked configurations. We have also taken advantage of the analytical expressions presented in our previous work to extended the analysis of the co- and cross-pol terms to FSS consisting of linearly transformed (rotated, scaled, misaligned) plates. Then, we have presented some examples in order to validate the present approach.  We have compared the circuit results with both numerical and experimental data extracted from the literature, as well as with data obtained with commercial full-wave simulator CST. This would be the case of an aperture-based circular polarizer and patch-based (single and stacked) structures formed by split-ring resonators and L-shaped dipoles.  Good agreement is observed for all the cases under study, even under extreme oblique incidence conditions.
In light of the results, the proposed ECA proves to be an efficient surrogate model that can be combined with traditional optimization techniques, metaheuristics or artificial intelligence algorithms for the practical design of FSS structures, polarizers, reflectarray cells and other microwave and photonic platforms, specifically when an accurate control of the cross-pol term is imperative.

\bibliographystyle{IEEEtran}
\bibliography{IEEEabrv,cross-polar}

\ifCLASSOPTIONcaptionsoff
  \newpage
\fi

\end{document}